\documentclass[sn-mathphys-num]{sn-jnl}

\usepackage{graphicx}%
\usepackage{amsmath,amssymb,amsfonts}%
\usepackage{amsthm}%
\usepackage{mathrsfs}%
\usepackage[title]{appendix}%
\usepackage[table]{xcolor}%
\usepackage{textcomp}%
\usepackage{manyfoot}%
\usepackage{booktabs}%
\usepackage{algorithm}%
\usepackage{algorithmicx}%
\usepackage{algpseudocode}%
\usepackage{listings}
\usepackage{url}
\usepackage{subcaption}
\usepackage{makecell}
\usepackage{booktabs}
\usepackage{multirow}
\usepackage{multicol}
\usepackage{array}
\usepackage{pdflscape}
\usepackage{hyperref}
\usepackage{pifont}
\usepackage{xspace}
\usepackage{adjustbox}
\usepackage[percent]{overpic}

\setcitestyle{super,open={},close={}}

\theoremstyle{thmstyleone}%
\theoremstyle{thmstyletwo}%

\theoremstyle{thmstylethree}%

\newcommand{\model}{\textsc{TrustTrade}\xspace}

\usepackage{pifont}

\begin{document}

\title[Article Title]{
\textsc{TrustTrade}: Human-Inspired Selective Consensus Reduces Decision Uncertainty in LLM Trading Agents
}


\author[1]{\fnm{Minghan} \sur{Li}}\email{mili4@meei.harvard.edu}

\author[1,2]{\fnm{Rachel} \sur{Gonsalves}}\email{rgonsalves@mba2027.hbs.edu}

\author[1]{\fnm{Weiyue} \sur{Li}}\email{weiyueli@fas.harvard.edu}

\author[3]{\fnm{Sunghoon} \sur{Yoon}}\email{shyoon@dgist.ac.kr}


\author*[1,4]{\fnm{Mengyu} \sur{Wang}}\email{Mengyu\_Wang@meei.harvard.edu}

\affil*[1]{\orgdiv{Harvard AI and Robotics Lab}, \orgname{Harvard University} 
}

\affil[2]{\orgdiv{Harvard Business School}, \orgname{Harvard University} 
}

\affil[3]{\orgname{Daegu Gyeongbuk Institute of Science and Technology} 
}

\affil[4]{\orgname{Kempner Institute for the Study of Natural and Artificial Intelligence, Harvard University}}


\abstract{

Large language models (LLMs) are increasingly deployed as autonomous agents in financial trading. However, they often exhibit a hazardous behavioral bias that we term uniform trust, whereby retrieved information is implicitly assumed to be factual and heterogeneous sources are treated as equally informative. This assumption stands in sharp contrast to human decision-making, which relies on selective filtering, cross-validation, and experience-driven weighting of information sources. As a result, LLM-based trading systems are particularly vulnerable to multi-source noise and misinformation, amplifying factual hallucinations and leading to unstable risk–return performance.
To bridge this behavioral gap, we introduce \model (Trust-Rectified Unified Selective Trader), a multi-agent selective consensus framework inspired by human epistemic heuristics. \model replaces uniform trust with cross-agent consistency by aggregating information from multiple independent LLM agents and dynamically weighting signals based on their semantic and numerical agreement. Consistent signals are prioritized, while divergent, weakly grounded, or temporally inconsistent inputs are selectively discounted. To further stabilize decision-making, \model incorporates deterministic temporal signals as reproducible anchors and a reflective memory mechanism that adapts risk preferences at test time without additional training. Together, these components suppress noise amplification and hallucination-driven volatility, yielding more stable and risk-aware trading behavior.
Across controlled backtesting in high-noise market environments (2024 Q1 and 2026 Q1), the proposed \model calibrates LLM trading behavior from extreme risk–return regimes toward a human-aligned, mid-risk/mid-return profile.
}

\keywords{large language models, autonomous trading agents, multi-agent consensus, factual hallucination,  decision uncertainty}



\maketitle

\label{sec:intro}





The rapid advancement of large language models (LLMs)~\cite{hurst2024gpt,team2023gemini,liu2024deepseek} has accelerated their deployment as autonomous agents~\cite{wang2024survey,ding2024survey} in financial decision-making~\cite{ingersoll1987theory}. Owing to their capacity to ingest and synthesize heterogeneous information, including market data, corporate fundamentals, news, and social sentiment, LLMs offer an appealing foundation for automated trading systems~\cite{lopezlira2025orderbook} operating at unprecedented scale and speed~\cite{ding2024survey,bai2024rlreview}. 

Recent research has consequently developed a range of advanced LLM-based trading frameworks~\cite{llmtrading2025moerouting,llmtrading2024drlportfoliochina,llmtrading2025decisioninformednn,llmtrading2024multimodalfoundationagent,zhang2024stockagent,cao2025chainofalpha}.
These approaches often leverage LLMs to incorporate chain-of-thought~\cite{wei2022chain}, external information retrieval, tool use, multi-agent collaboration~\cite{tran2025multi} or reinforcement learning~\cite{sutton1998reinforcement} for portfolio optimization~\cite{llmtrading2024drlportfoliochina,coriat2025harlf}, market sentiment analysis~\cite{unnik2024sentiment}, cross-modal information integration~\cite{xiao2025tradingagents}, and end-to-end trading decision generation~\cite{llmtrading2025decisioninformednn,li2025r}. Overall, these cutting-edge techniques extend model capabilities in both simulated and realistic market environments.

However, in quantitative trading settings\cite{mandelbrot1963variation,engle1982autoregressive,diebold2014network}, the limitations of LLM-based approaches have become increasingly evident. Because LLM reasoning is inherently stochastic and highly sensitive to how information is formulated and combined, trading decisions can be overly responsive to multi-source noise and minor input perturbations, resulting in unstable and poorly reproducible outcomes.
To mitigate uncertainty~\cite{brock1997rational} and non-stationarity~\cite{brock1998heterogeneous} in financial markets, 
recent approaches introduce structural constraints through multi-agent strategy search and evolutionary modeling~\cite{yun2025quantevolve,han2026quantaalpha}, or through logic-oriented representations of market narratives using semantic and temporal alignment~\cite{guo2026meme}; however, these methods primarily focus on strategy generation and market structure characterization.

In contrast, we argue that decision uncertainty in LLM-based trading arises from a more fundamental limitation: a systematic bias in how information is trusted and integrated. Recent studies show that under open-ended retrieval and online search, LLM agents are prone to factual hallucinations and spurious correlations~\cite{benhenda2025finrldeepseek,lee2025bias,lin2025llm}. In trading settings, such failures extend beyond semantic errors and can directly influence portfolio decisions, amplifying volatility and drawdowns.

Crucially, this vulnerability is not driven by insufficient model capacity but by an implicit uniform-trust assumption in many existing approaches, which treat information retrieved or generated by LLM agents as factual and equally reliable~\cite{xiao2025tradingagents,swanson2024virtual,piatti2024cooperate}. This assumption ignores substantial variation in source quality, reliability, and temporal relevance. In high-noise, temporally dependent, and potentially adversarial financial environments, even minor distortions can therefore be magnified into disproportionate risk exposure. Our empirical results indicate that this limitation reflects a behavioral mismatch between LLM agents and human annotators, rather than a purely technical shortcoming.

In our empirical study, human annotators follow a markedly different strategy in financial decision-making~\cite{preis2013quantifying}. Rather than weighting information uniformly, human annotators rely on selective filtering~\cite{coval2001information}: they prioritize signals that are temporally coherent, repeatedly corroborated, and historically reliable, while down-weighting narrative-driven, emotional~\cite{tuckett2011emotions}, noisy, or weakly grounded inputs. This process is tightly coupled with memory~\cite{kahneman2011thinking,froot1992explaining} (such as past experiences, prior outcomes, and long-term context), which stabilizes decision-making and mitigates overreaction to transient market narratives.

To bridge this behavioral gap, we introduce \model (Trust-Rectified Unified Selective Trader), a multi-agent selective consensus framework inspired by human epistemic heuristics. Instead of relying on a single LLM agent, our approach deploys multiple independent agents to collect and interpret information in parallel. The framework operationalizes the principle of invariance of truth: objective and reliable signals should remain consistent across independent reasoning paths. By quantifying cross-agent agreement in both semantic and numerical space, we derive dynamic credibility scores that govern information weighting. High-consensus signals are prioritized, while divergent, weakly grounded, or temporally inconsistent inputs, including those implicitly encoding look-ahead bias, are selectively discounted. 

Beyond consensus filtering, \model incorporates a deterministic temporal signal module that compresses raw price dynamics and market states into reproducible and auditable time-series indicators, including trend, momentum, volatility, drawdown, and risk exposure. These signals serve as stable anchors for downstream decision-making, constraining the influence of unreliable textual evidence and providing a consistent temporal reference across trading horizons. As a result, strategy updates become less sensitive to spurious information and less prone to abrupt fluctuations driven by factual hallucinations.

Finally, \model integrates a memory bank with short- and long-term decision reflection mechanisms to enable test-time adaptation without additional training. The system continuously records historical trading contexts, including supporting evidence, consensus scores, temporal signals, executed actions, and confidence levels, and performs retrospective evaluations based on realized return–risk outcomes. Through this process, the model progressively calibrates its risk preferences and information-weighting strategy.

We evaluate \model through controlled A/B experiments against standard LLM agents and human annotators in high-noise market simulations and real-world markets. The results demonstrate that \model fundamentally reshapes LLM behavior: maximum drawdown is substantially reduced, risk–return trade-offs shift toward the human-preferred regime, and performance becomes more stable. Importantly, these improvements arise not from increased model capacity, but from behavioral alignment.


    
    


\section*{Results}\label{sec:results}


\begin{figure}[!htbp]
    \centering
    \begin{overpic}[width=0.96\textwidth]{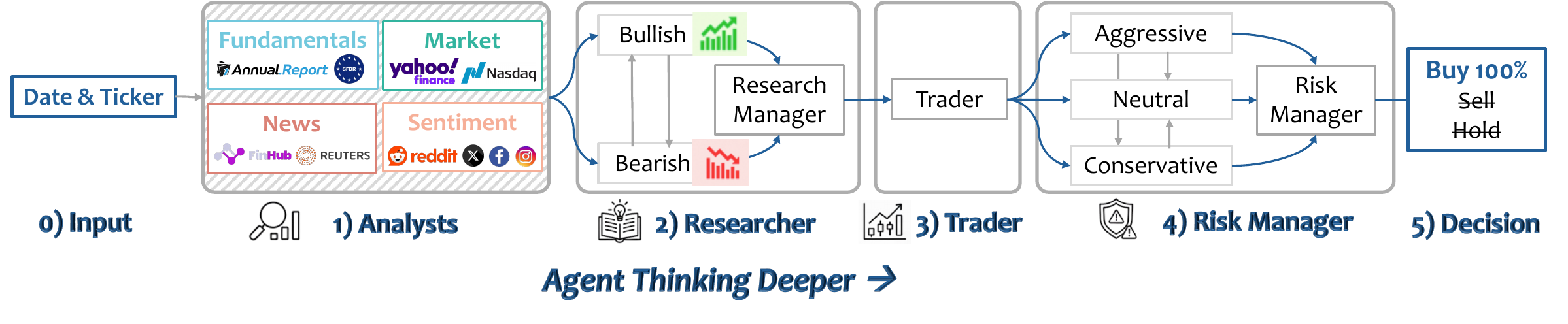}
        \put(-3,18){\color{blue} \textbf{a}}\label{fig:ta}
    \end{overpic}
    \vfill
    \begin{overpic}[width=0.96\textwidth]{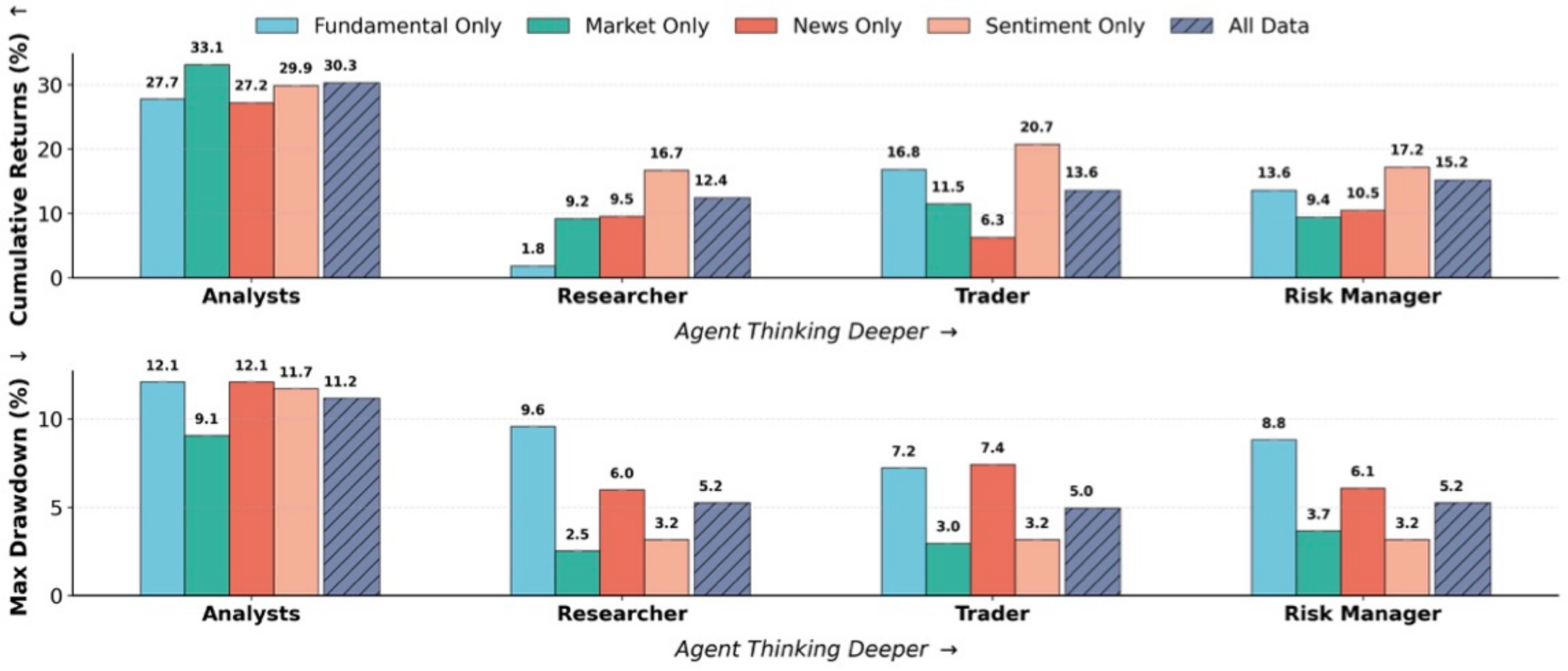}
        \put(-3,40){\color{blue} \textbf{b}}
        \put(-3,18){\color{blue} \textbf{c}}
    \end{overpic}
    \vfill
    \begin{overpic}[width=0.96\textwidth]{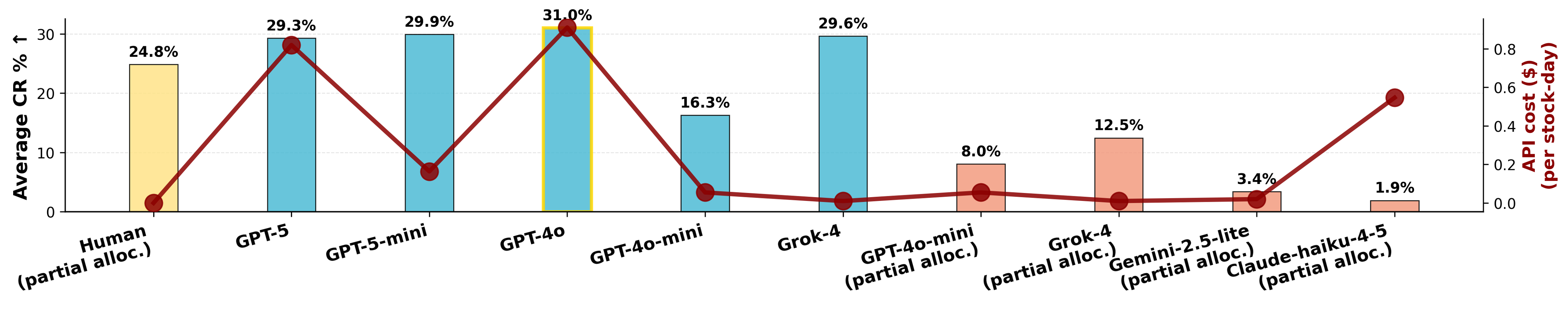}
        \put(-3,17){\color{blue} \textbf{d}}
    \end{overpic}
    \vfill
    \begin{overpic}[width=0.96\textwidth]{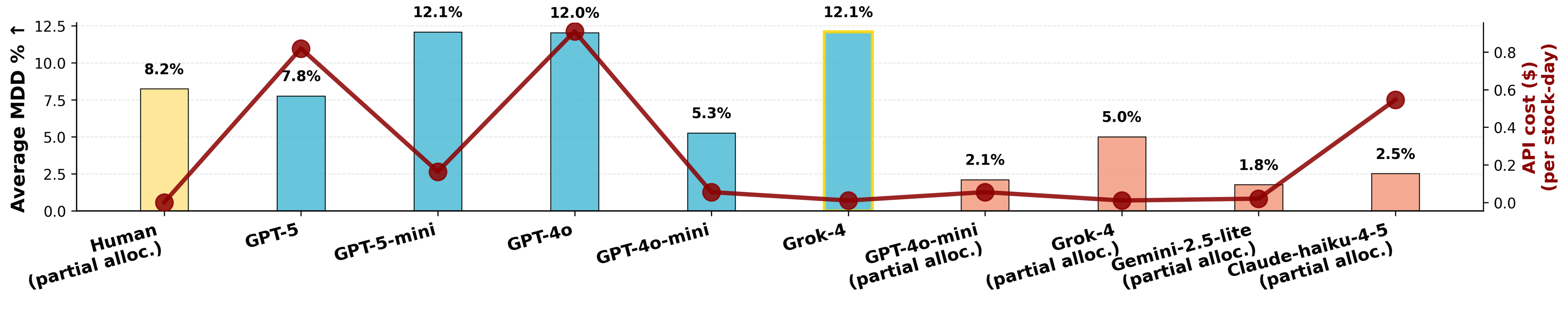}
        \put(-3,17){\color{blue} \textbf{e}}
    \end{overpic}
    \vfill
    \begin{overpic}[width=0.32\textwidth]{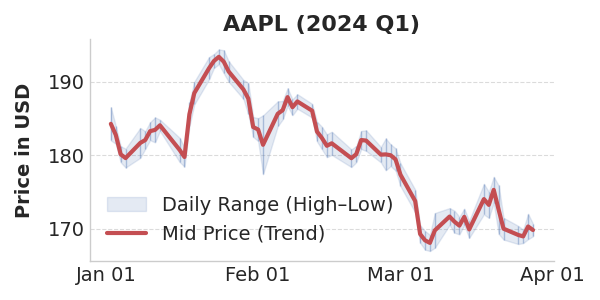}
        \put(2,43){\color{blue} \textbf{f}}
    \end{overpic}
    \begin{overpic}[width=0.32\textwidth]{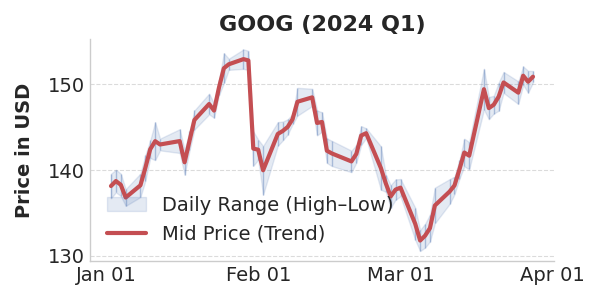}
        \put(2,43){\color{blue} \textbf{g}}
    \end{overpic}
    \begin{overpic}[width=0.32\textwidth]{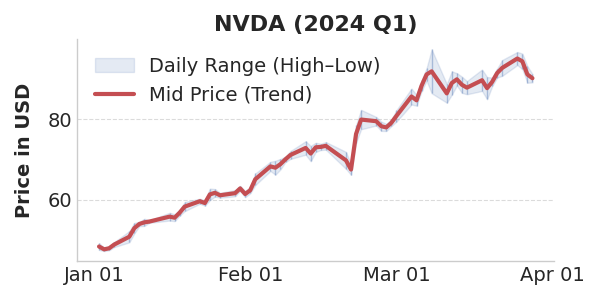}
        \put(2,43){\color{blue} \textbf{h}}
    \end{overpic}
    \caption{\small 
\textbf{Diagnosing instability in LLM-based trading agents (with GPT-4o-mini).} 
This figure illustrates how data sources, reasoning depth, and allocation regimes jointly contribute to the instability and risk–return profiles of LLM trading agents.
\textbf{\color{blue} a,} Schematic of a standard LLM-based trading pipeline~\cite{xiao2025tradingagents} with increasing reasoning depth, progressing from Analysts to Researcher, Trader, and Risk Manager.
\textbf{\color{blue} b-c,} Cumulative returns and maximum drawdown across LLM agents under different data-source and reasoning-depth ablations. Return and risk vary across data sources, transitioning from high-risk/high-return to more balanced trade-offs with increasing reasoning depth.
\textbf{\color{blue} d-e,} Risk--return heterogeneity across human and LLM trading agents under different allocation regimes. Average cumulative return (CR) and maximum drawdown (MDD) across stocks, with API cost per stock-day shown on the right axis. Full-allocation agents earn higher returns but suffer larger drawdowns than partial-allocation agents and human annotators.
\textbf{\color{blue} f-h}, Stock price dynamics during 2024-Q1. Note that all reported results are averaged over these three stocks (AAPL, GOOG, NVDA) during 2024-Q1.
}\label{fig:study1}
\end{figure}

\subsection*{Study I: Diagnosing Instability in LLM-Based Trading Agents}
Across all three market regimes examined in 2024-Q1 (Fig.~\ref{fig:study1}~{\color{blue}\textbf{f-h}}), we observe pronounced heterogeneity in trading behavior among LLM agents, whose structure is showed in Fig.~\ref{fig:study1}~{\color{blue}\textbf{a}}. Even when exposed to identical market conditions, different agents frequently arrive at divergent trading decisions, leading to substantial variation in cumulative returns, drawdowns, and risk-adjusted performance (Fig.~\ref{fig:study1}~{\color{blue}\textbf{b-e}}). This heterogeneity persists across reasoning depth, agent architectures, and the composition of accessible information sources, indicating that trading outcomes are not determined solely by market dynamics but are strongly shaped by model-specific decision processes.

\noindent\textbf{Reasoning depth of LLM agents.}
We examined how trading performance changes as LLM agents operate with increasing reasoning depth (Fig.~\ref{fig:study1}~{\color{blue}\textbf{b-c}}). The results reveal a clear stage-wise structure rather than a gradual or monotonic improvement. Advancing from the Analyst to the Trader stage leads to substantial gains in decision quality, reflected by markedly reduced maximum drawdowns. This transition captures nearly all performance benefits associated with deeper reasoning, indicating that the integration of analysis and execution constitutes the critical bottleneck for effective trading decisions.
In contrast, adding a dedicated Risk Manager stage does not produce systematic benefits. Its behavior largely overlaps with that of the Trader stage and fails to consistently improve either returns or risk control. On the basis of these results, the Risk Manager stage is excluded from subsequent experiments, as it introduces additional complexity without measurable gains in decision quality.

\noindent\textbf{Non-additive effects of information modalities.}
Information sources have non-additive effects (Fig.~\ref{fig:study1}~{\color{blue}\textbf{b,c}}). At the Analyst stage, market-only signals perform best (33.1\% CR; 9.1\% MDD), while adding fundamentals/news/sentiment reduces returns and increases drawdown, indicating that naive multi-source aggregation injects noise. At the Trader stage, sentiment-only appears strong but is prone to leakage because it is retrieved via online search; combining sources still fails to improve performance (13.6\% return; 5.0\% MDD), suggesting limited ability to filter unreliable evidence.


\noindent\textbf{Systematic bias across LLM agent variants.}
Fig.~\ref{fig:study1}~{\color{blue}\textbf{d–e}} reveals clear heterogeneity in risk–return behavior across LLM agents and human annotators.
Under full allocation, large LLMs (GPT-5, GPT-5-mini, GPT-4o, and Grok-4) achieve the highest cumulative returns ($\approx$30\% on average), but consistently incur large drawdowns ($\approx12$\% MDD), indicating aggressive, high-confidence trading with limited risk control. In contrast, GPT-4o-mini exhibits substantially lower returns but markedly reduced drawdowns, despite operating under the same allocation constraint, suggesting more conservative timing and weaker signal amplification. 
%
Under partial allocation, GPT-4o-mini and Grok-4 shift toward a risk-controlled regime, with returns decreasing from 29.6\% to 12.5\% and drawdowns reduced from 12.1\% to 5.0\%. Smaller models (Gemini-2.5-lite and Claude-Haiku-4.5) consistently produce low returns ($<$4\%) with minimal drawdowns ($<$2.5\%), reflecting strongly risk-averse behavior. 
%
Overall, these results show a systematic separation in behavior: larger LLMs favor high-reward strategies with elevated downside risk, whereas smaller models and human annotators prioritize drawdown control, highlighting a fundamental gap between LLM-driven and human behavior.

\subsection*{Study II: Characterizing Human Stabilization Mechanisms}

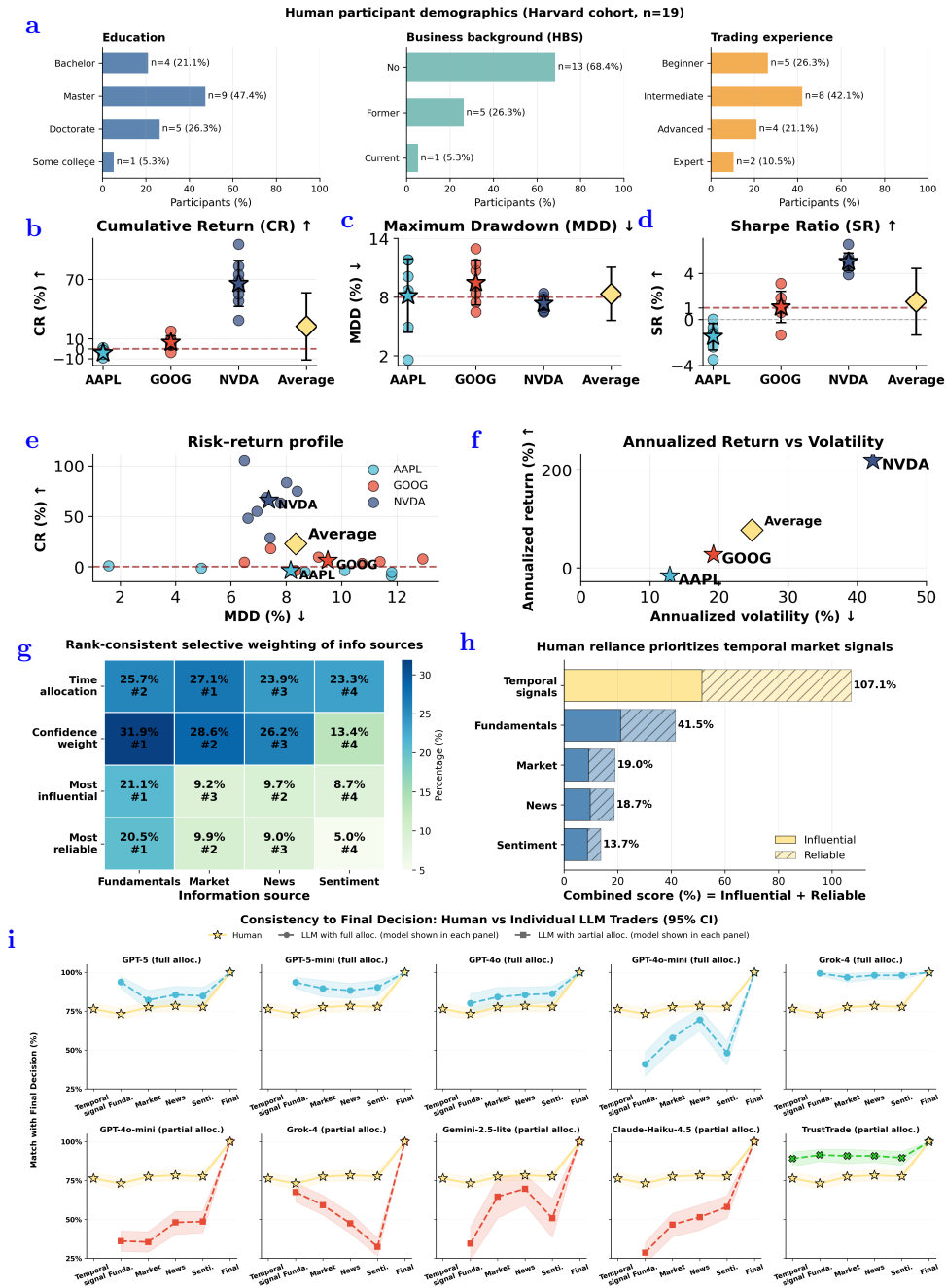
\begin{figure}[!htbp]
    \begin{overpic}[width=0.95\textwidth]{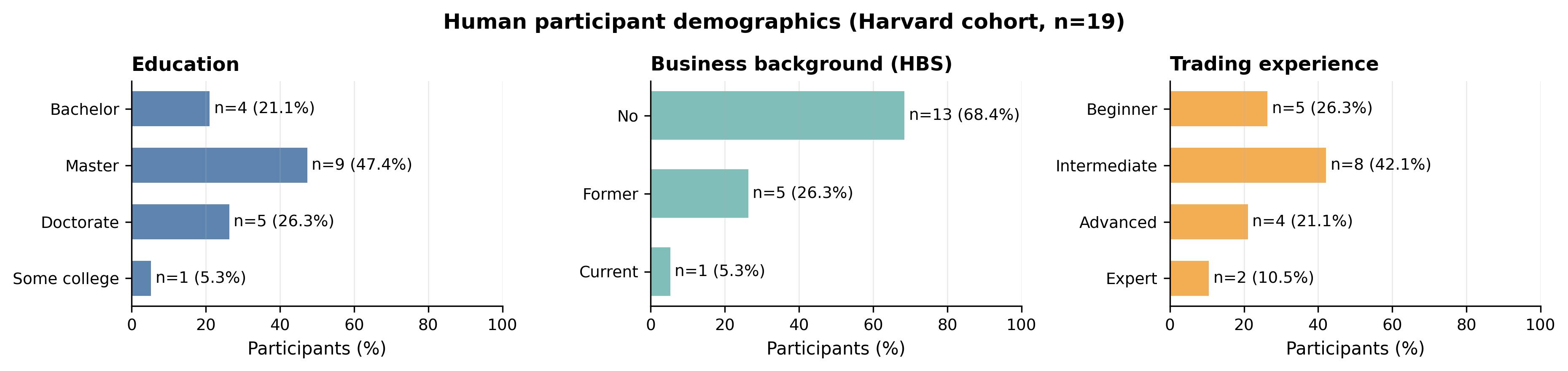}
        \put(0,20){{\color{blue}\textbf{a}}}
    \end{overpic}
    \vfill
    \begin{overpic}[width=0.96\textwidth]{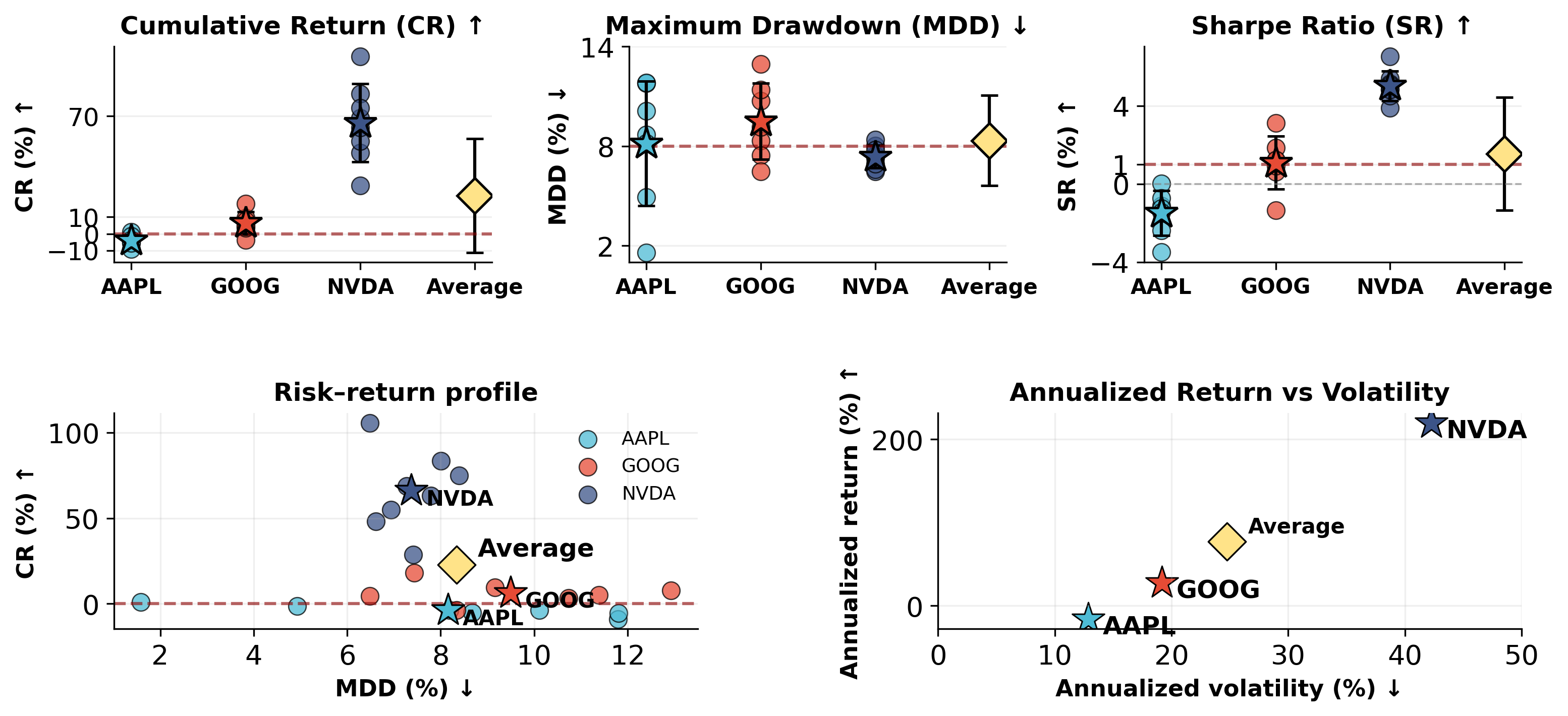}
        \put(0,43){{\color{blue}\textbf{b}}}
        \put(34,44){{\color{blue}\textbf{c}}}
        \put(66,44){{\color{blue}\textbf{d}}}
        \put(0,20){{\color{blue}\textbf{e}}}
        \put(48,20){{\color{blue}\textbf{f}}}
    \end{overpic}
    \vfill
    \begin{overpic}[width=0.44\textwidth]{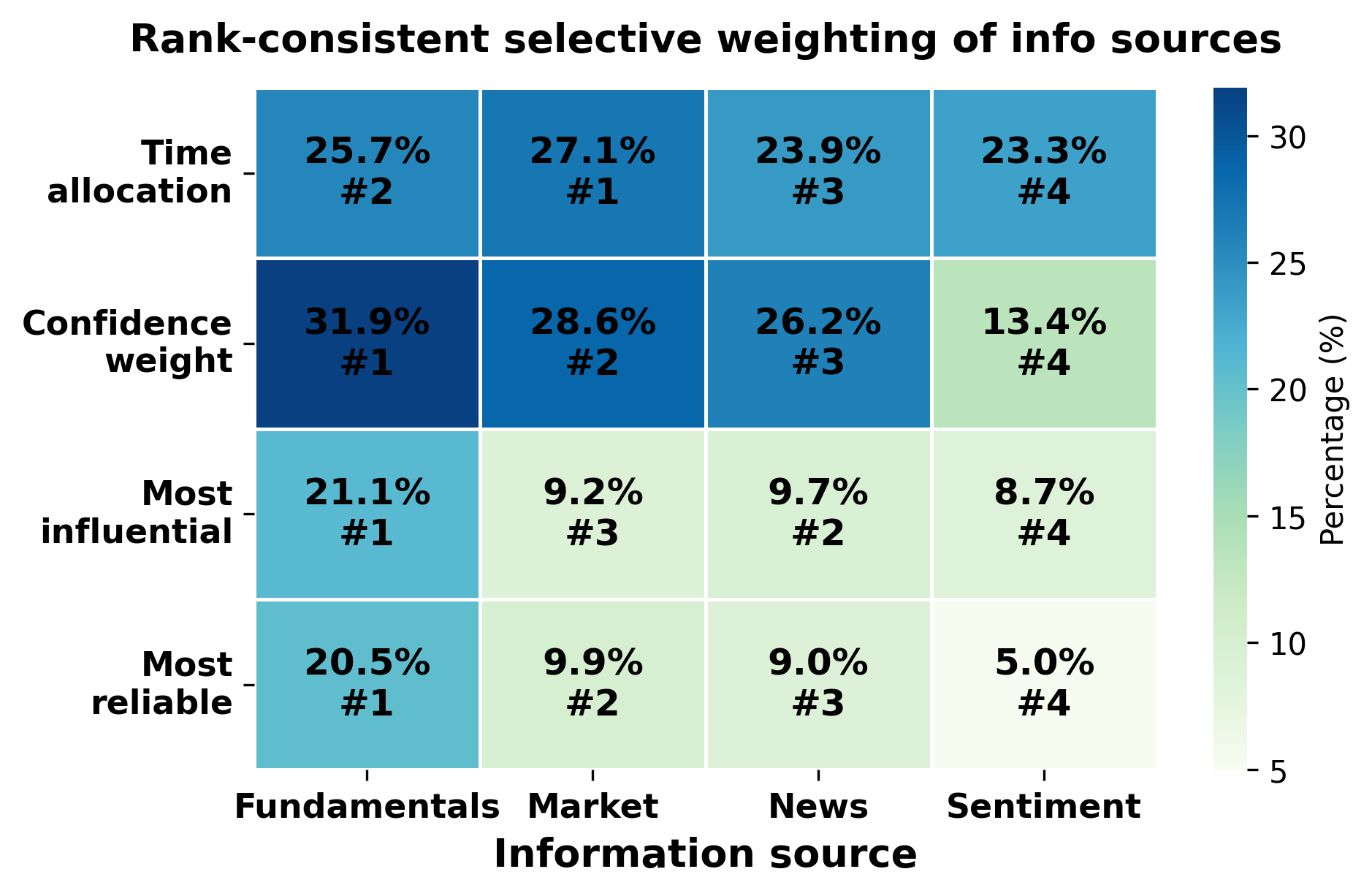}
        \put(-2,60){\color{blue} \textbf{g}}
    \end{overpic}
    \begin{overpic}[width=0.45\textwidth]{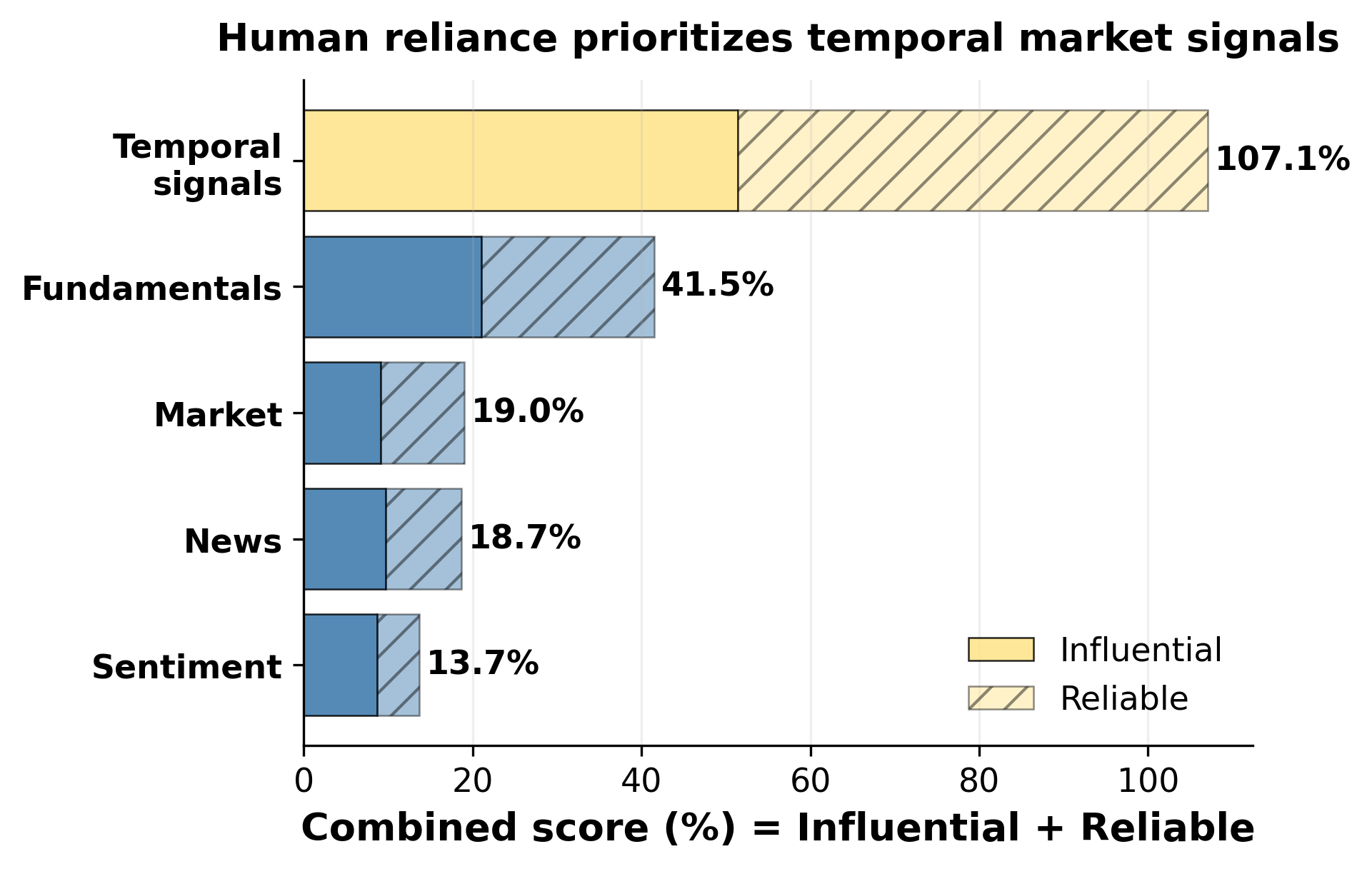}
        \put(-2,60){\color{blue} \textbf{h}}
    \end{overpic}
    \vfill
    \begin{overpic}[width=0.95\textwidth]{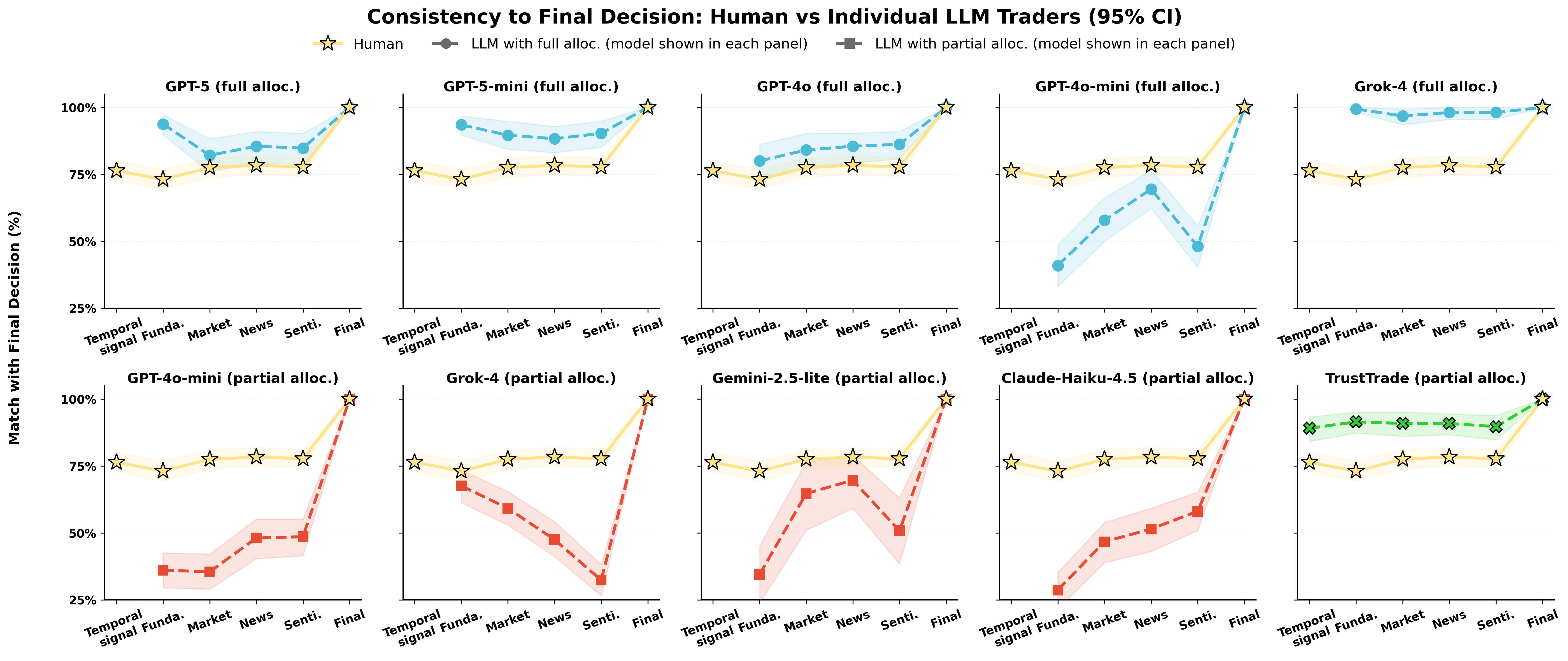}
         \put(-2,38){\color{blue} \textbf{i}}
    \end{overpic}
    \caption{\footnotesize 
    \textbf{Behavioral signatures of human trading.}
     \textbf{\color{blue} a,} Demographic profile of human annotators (n=19). 
     \textbf{\color{blue} b–d,} Human trading outcomes across stocks, showing moderate cumulative returns, tightly controlled drawdowns, and near-neutral Sharpe ratios, with consistent risk exposure across assets. 
     \textbf{\color{blue} e-f,} Risk–return and volatility profiles illustrate that human exhibits substantially greater dispersion across assets. 
    \textbf{\color{blue} g-h,} Selective information weighting by human annotators: time allocation, confidence weighting and post-hoc influence/reliability are aligned, and the combined influence--reliability score emphasizes temporally grounded signals (price trends and market indices) over narrative-driven inputs (news and sentiment). 
    \textbf{\color{blue} i,} Decision convergence across sequential information stages for human annotators and nine LLM-based traders under full- and partial-allocation settings. Human annotators show consistently high convergence to the final action across stages, whereas LLM traders exhibit lower and more variable convergence.
    }
    \label{fig: human_analysis}
\end{figure}


Human trading behaviour was studied using an online, stepwise simulation with structured logging. Participants progressed through a fixed sequence of information stages (temporal signals, fundamentals, market indicators, news and sentiment) and provided stage-wise actions, confidence ratings and rationales. We additionally recorded time spent on each information source and post-hoc judgments of influence and reliability. These signals yield reproducible behavioural traces that enable a direct comparison with sequential LLM decision pipelines (Fig.~\ref{fig: human_analysis}). Nineteen human annotators were recruited (students or faculty members at Harvard University), whose personas are shown in Fig.~\ref{fig: human_analysis}~{\color{blue}\textbf{a}}.

\noindent\textbf{Human trading exhibits stable risk--return integration.}
Human annotators exhibit moderate returns with consistently controlled risk across assets in Fig.~\ref{fig: human_analysis}~{\color{blue}\textbf{b–d}}. While cumulative returns vary by asset, highest for NVDA and lowest for AAPL, performance remains stable and avoids extreme outcomes. Maximum drawdowns are relatively contained in all cases, indicating a strong preference for capital preservation over aggressive return maximization. In the risk–return space of Fig.~\ref{fig: human_analysis}~{\color{blue}\textbf{e}}, human strategies cluster within a mid-risk, mid-return regime, avoiding both high-volatility/high-return and low-risk/low-return extremes. Consistent with this behavior, the annualized return–volatility analysis in Fig.~\ref{fig: human_analysis}~{\color{blue}\textbf{f}} shows that human annotators systematically accept lower returns in exchange for reduced volatility, in contrast to strategies that aggressively exploit high-volatility opportunities.

\noindent\textbf{Empirical evidence of data filtering and temporal signals.}
Human annotators exhibit a highly structured and selective pattern of information use. In Fig.~\ref{fig: human_analysis}~{\color{blue}\textbf{g}}, we separately quantify (i) the time that human annotators spend on each information source, (ii) their confidence weight assigned to that source (0-100), and (iii) post-hoc selections of the most influential and most reliable sources.
The proportion of time allocation, confidence weighting, and post-hoc judgments of influence and reliability are strongly rank-consistent across information sources, with fundamentals and market signals consistently prioritized over news and sentiment. 
Together, these measures point to an implicit filtering strategy in human information use, in which attention and confidence co-vary with perceived utility rather than being allocated independently across sources.
Fig.~\ref{fig: human_analysis}~{\color{blue}\textbf{h}} reveals a second, complementary signature of human stabilization: a marked preference for temporally anchored evidence. When influence and reliability are aggregated, temporal signals dominate (combined score, 107.1\%), exceeding fundamentals, market signals, news and sentiment by a wide margin. Thus, even in the presence of narrative inputs, human judgements remain grounded in verifiable price dynamics and benchmark movements, with news and sentiment treated as ancillary and comparatively unreliable.
%

\noindent\textbf{Decision convergence distinguishes human and LLM trading decisions.} 
Decision convergence reveals fundamental behavioral differences between human annotators and LLM traders.
Fig.~\ref{fig: human_analysis}~{\color{blue}\textbf{i}} shows the consistency between intermediate decisions and the final trading action across sequential information stages. Human annotators exhibit uniformly high convergence throughout the decision pipeline, indicating that early-stage judgments are largely preserved and incrementally refined rather than overwritten. This pattern suggests stable internal representations supported by selective data filtering and historical memory.
In contrast, LLM traders show pronounced stage-wise inconsistency. Across models and allocation regimes, intermediate decisions often diverge sharply from the final action, with convergence occurring abruptly only at the final stage. This behavior indicates reactive decision revision rather than cumulative reasoning, consistent with limited temporal memory and weak commitment to earlier evidence.
Partial-allocation strategies moderately improve convergence for some models, particularly GPT-4o-mini and Grok-4, but substantial instability remains across intermediate stages. Overall, these results demonstrate that human annotator behavior is characterized by early stabilization and memory-consistent integration, whereas current LLM traders rely on late-stage aggregation, highlighting a core behavioral gap in sequential decision-making.

\begin{figure}[!htbp]
    \centering
    \begin{overpic}[width=0.92\textwidth]{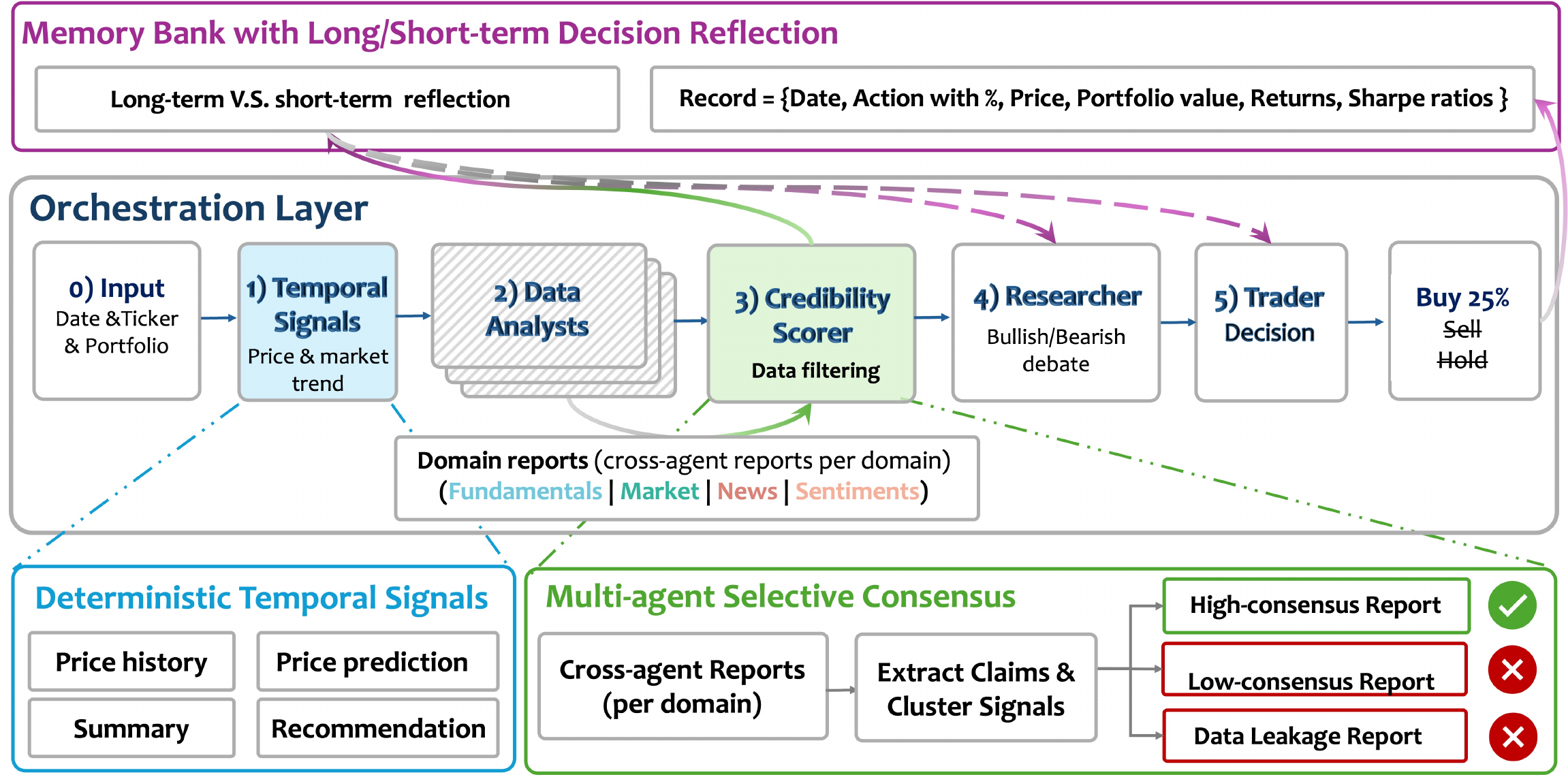}\label{fig:ta_ours}  
        \put(-3,48){\color{blue} \textbf{a}}
    \end{overpic}
    \vfill
    \begin{overpic}[width=0.48\textwidth]{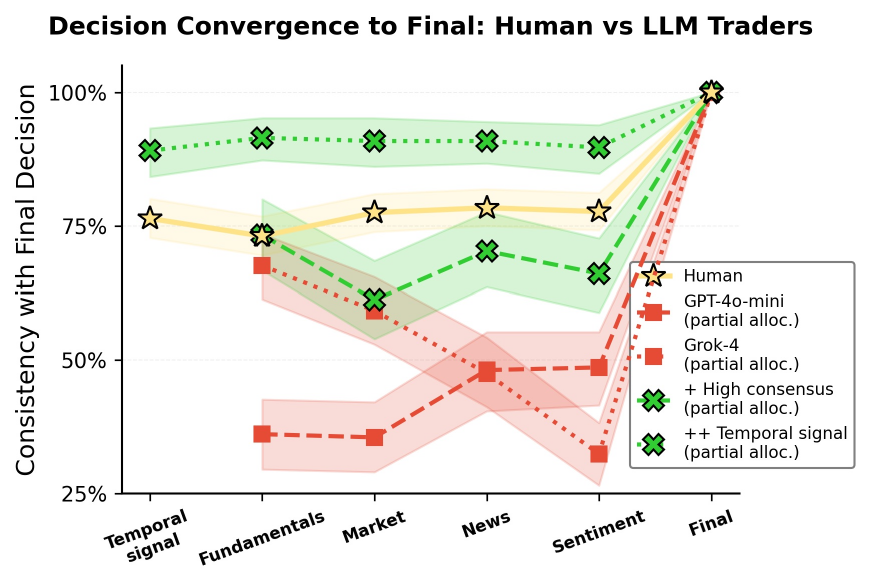}  
        \put(-3,62){\color{blue} \textbf{b}}
    \end{overpic}
    \begin{overpic}[width=0.48\textwidth]{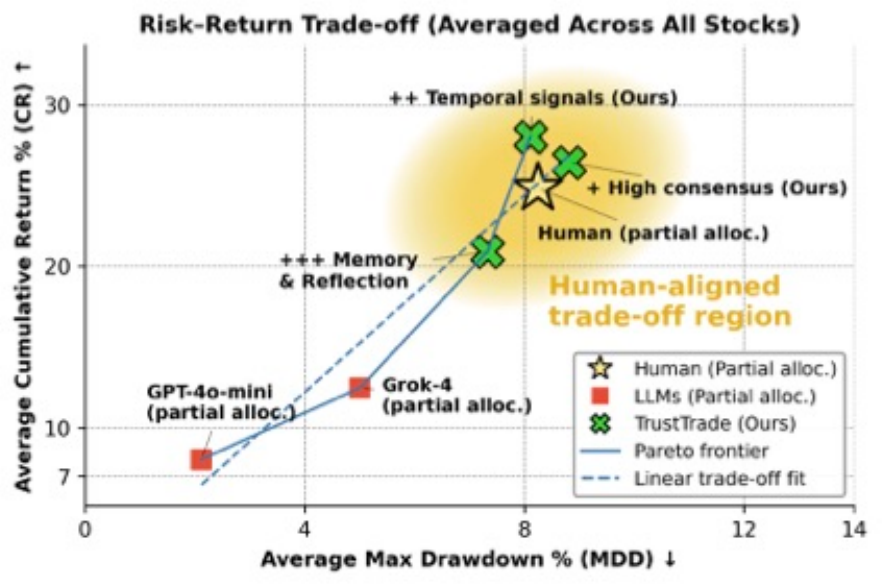}  
        \put(1,62){\color{blue} \textbf{c}}
    \end{overpic}
    \vfill
    \begin{minipage}{0.95\textwidth}
        \centering
        \makebox[0pt][l]{\hspace*{-2mm}\raisebox{2.2cm}[0pt][0pt]{\color{blue}\textbf{d}}}
        \centering
{\footnotesize
\setlength{\tabcolsep}{5pt}
\renewcommand{\arraystretch}{1.05}
\rowcolors{3}{black!3}{white}
\begin{tabular}{lll}
\multicolumn{3}{c}{\textbf{Temporal Signals Summary Table}} \\
\midrule
\textbf{Temporal Signal} & \textbf{Value} & \textbf{Notes / Implications} \\
\midrule
Time (Last Close)     & 2024-01-03   & Date corresponding to the current (closing) price \\
Current Price        & \$48.14      & Closing price on 2024-01-03 \\
1-Week Return        & -2.25\%      & Short-term downtrend; negative momentum \\
1-Month Return       & +5.85\%      & Medium-term uptrend holding \\
3-Month Return       & +7.58\%      & Strong uptrend; consistent gains \\
6-Month Return       & +15.03\%     & Building momentum from lows \\
1-Year Return        & +192.11\%    & Long-term bullish dominance \\
\midrule
1-Week Volatility    & 1.45\%       & Low risk, but limited upside \\
1-Month Volatility   & 1.78\%       & Moderate; supports trend continuation if broken \\
3-Month Volatility   & 2.11\%       & Increasing; watch for pullbacks \\
6-Month Volatility   & 2.28\%       & Higher risk in extended holds \\
\midrule
Next-Day Prediction  & DOWN 1.13\%  & Bearish signal; confidence 100\% \\
Key Support Level    & \$47.57      & 1.2\% downside risk \\
Key Resistance Level & \$49.97      & 3.8\% upside to break \\
Trend Alignment Score& 60 / 100     & Mixed (short-term weak, long-term strong) \\
\midrule
\multicolumn{3}{l}{\textbf{Temporal-signal-driven proposal:} \textbf{HOLD 0\%}} \\
\bottomrule
\end{tabular}
}
    \end{minipage}
    \vspace{1mm}
    \caption{\footnotesize
    \textbf{Human-aligned trading behavior induced by multi-agent consensus filtering, temporal signals, and memory bank with long/short-term decision reflection.}
    \textbf{\color{blue} a,} Overview of the proposed \model framework: multiple agents collect information from diverse sources, a credibility scorer filters for high-consensus evidence, and the resulting decision is used to update a memory bank with both short- and long-term reflections.
    \textbf{\color{blue} b,} Decision convergence across sequential stages, comparing human annotators, baseline LLM traders, and the proposed \model.
    \textbf{\color{blue} c,} Risk--return trade-off averaged across stocks; the shaded ellipse marks the human-aligned preference region defined by the standard errors of human-annotator cumulative return (CR) and maximum drawdown (MDD).
    \textbf{\color{blue} d,} Newly introduced temporal-signal summary reporting deterministic, price-derived trends and indicators to reduce hallucination and improve decision reliability.
    }
    \label{fig:study3}
\end{figure}

\begin{figure}[!htbp]
    \centering
    \begin{overpic}[width=0.48\textwidth]{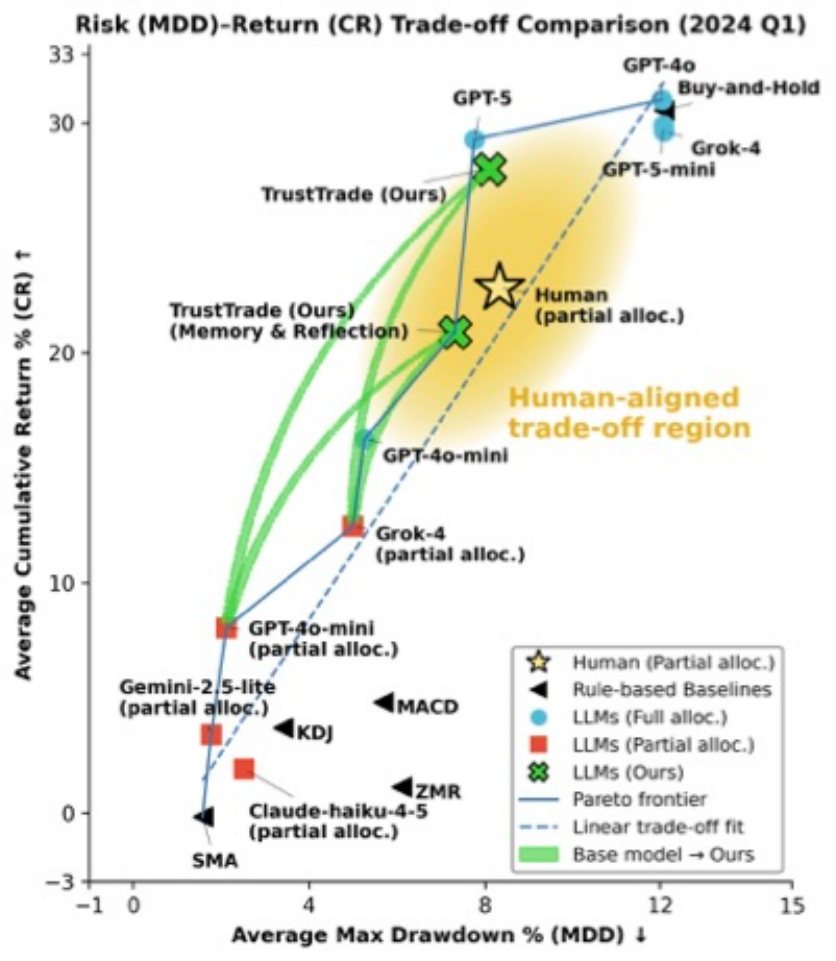}
        \put(-3,95){\textbf{\color{blue} a}}
    \end{overpic}
    \begin{overpic}[width=0.48\textwidth]{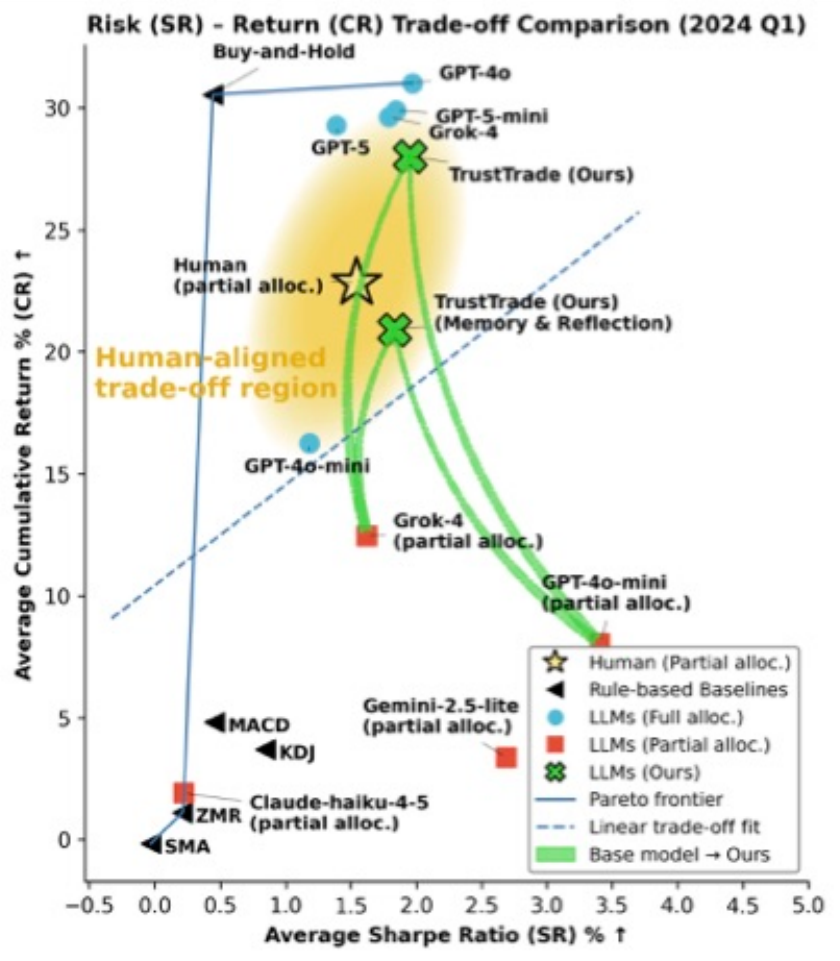}
        \put(-3,95){\textbf{\color{blue} b}}
    \end{overpic}
    \begin{overpic}[width=0.48\textwidth]{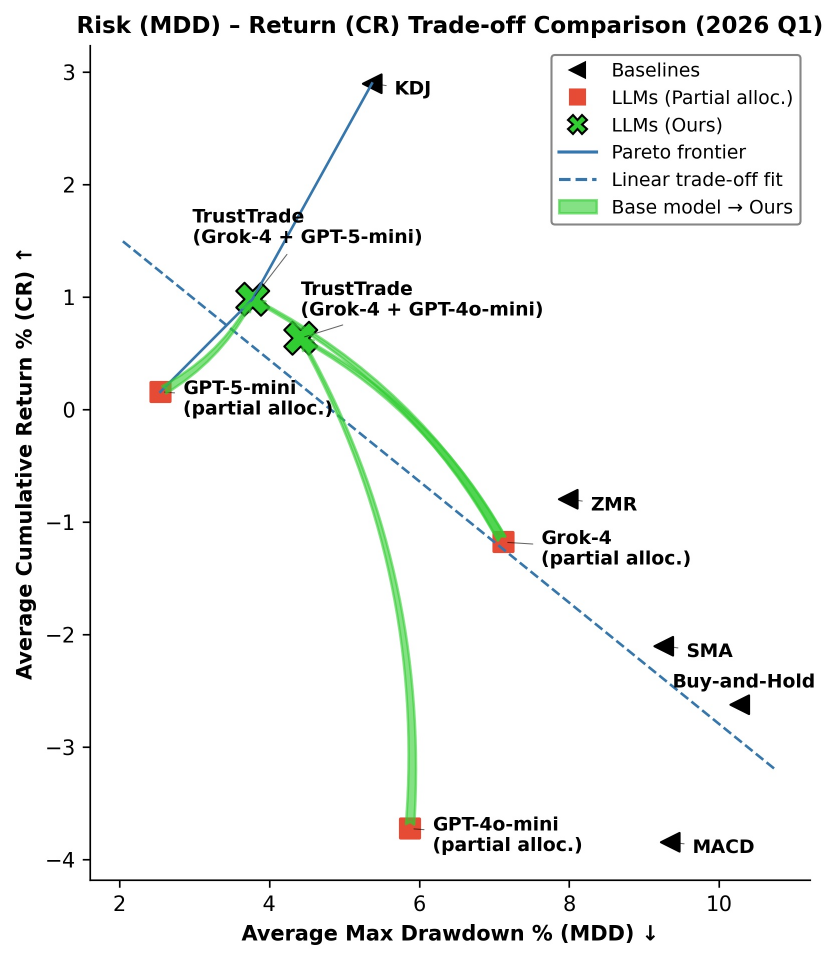}
        \put(-3,95){\textbf{\color{blue} c}}
    \end{overpic}
    \begin{overpic}[width=0.48\textwidth]{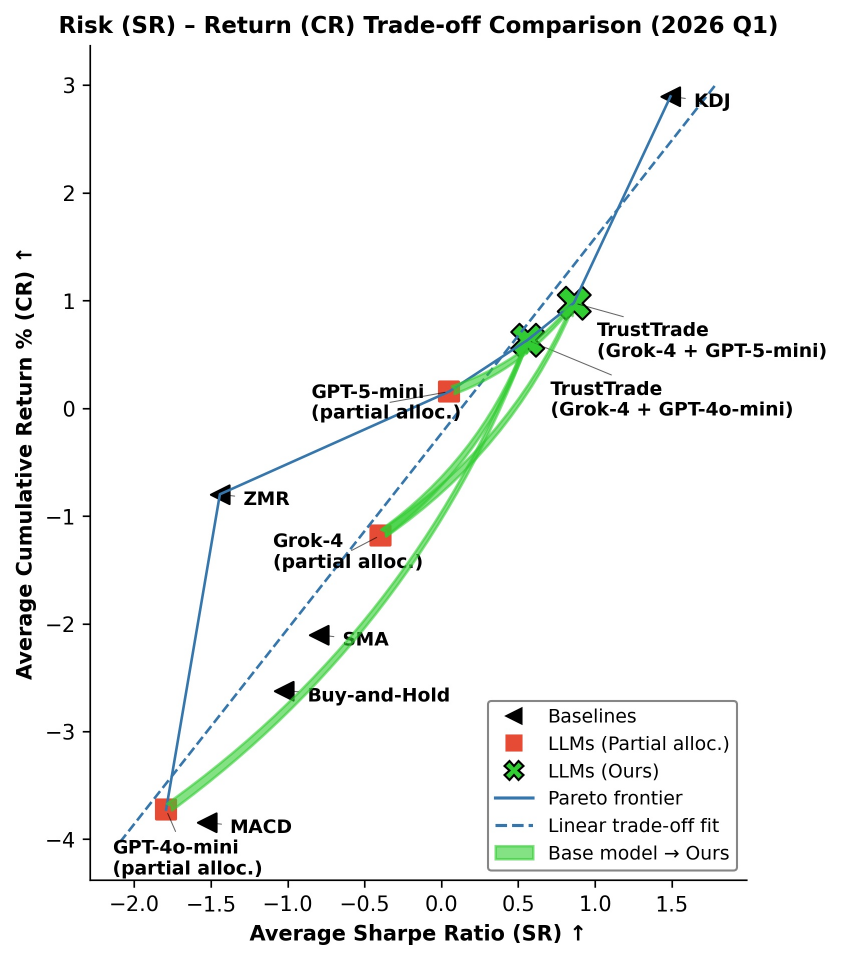}
        \put(-3,95){\textbf{\color{blue} d}}
    \end{overpic}
    \caption{\footnotesize 
    \textbf{Comprehensive backtesting comparison across all baselines over the 2024 Q1 and 2026 Q1 trading period respectively.}
    \textbf{\color{blue} a \& b,} Risk--return trade-off averaged across NVDA, AAPL, and GOOG in 2024 Q1. Each point represents an agent configuration, plotted by average cumulative return (CR) and average maximum drawdown (MDD) / average Sharpe Ratio (SR). Human annotators (yellow star) define a human-aligned risk--return preference region (shaded ellipse). Full-allocation LLM agents (blue dots) achieve high returns but incur substantial drawdowns, whereas partial-allocation LLMs (red squares) reduce risk at the cost of diminished returns. Our \model builds on GPT-4o-mini and Grok-4 under partial allocation, and we compare two variants: without memory and reflection, the method achieves higher returns than human annotators at comparable maximum drawdown (approaching GPT-5 performance), while adding memory and reflection slightly reduces returns but further lowers risk. The Pareto frontier and linear trade-off fit are shown for reference.
    \textbf{\color{blue} c \& d,} Day-by-day backtesting risk--return trade-off across NVDA, AAPL, and GOOG during 2026 Q1, comparing rule-based baselines, single-LLM traders, and our multi-LLM framework. Our \model achieves a substantially improved return--risk balance, with higher returns and lower risk than the comparison methods.
    }
    \label{fig:main}
\end{figure}

\begin{figure}[!htbp]
    \centering    
    \begin{overpic}[width=0.7\textwidth]{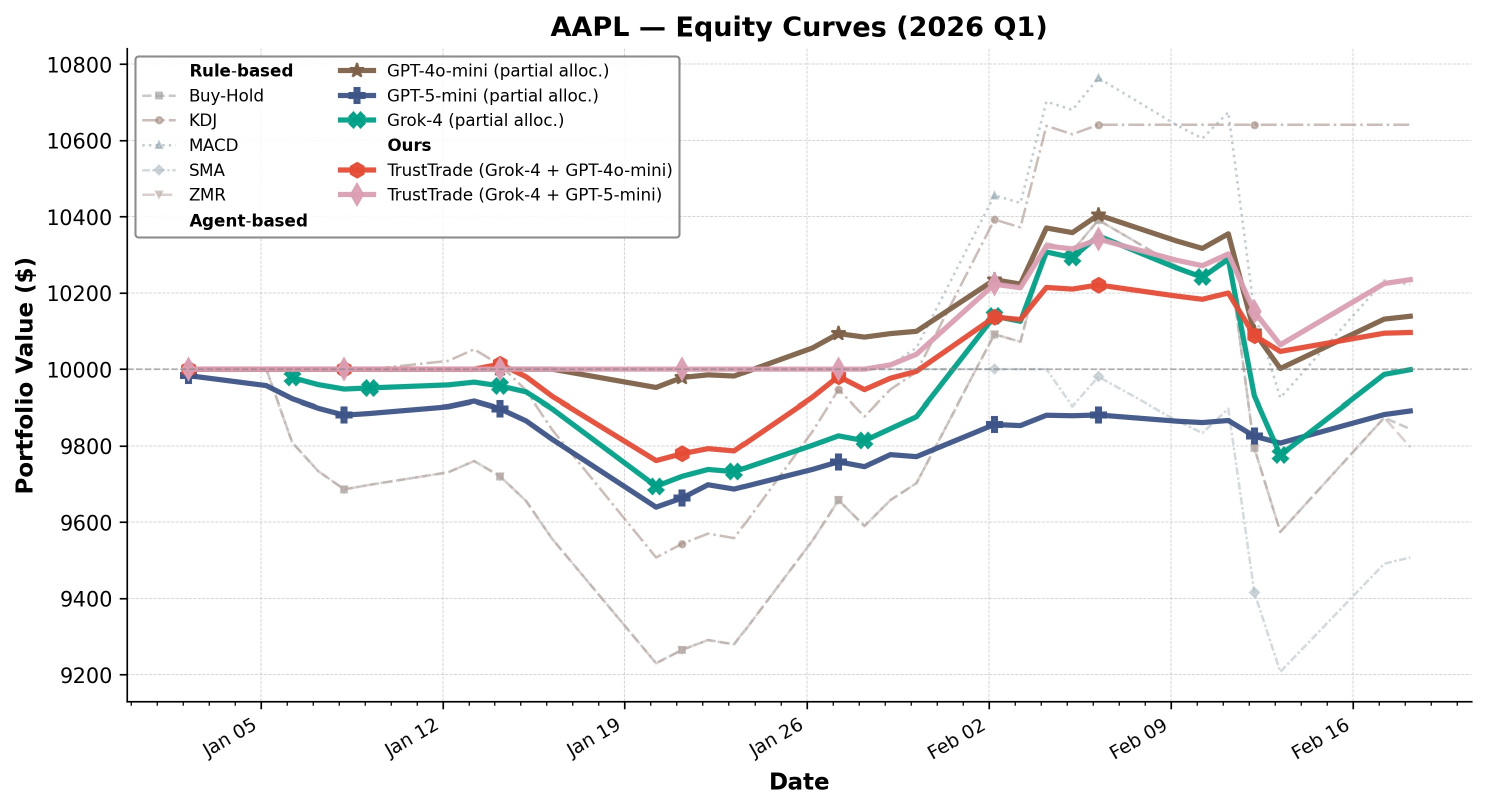}
        \put(-3,48){\textbf{\color{blue} a}}
    \end{overpic}
    \begin{overpic}[width=0.7\textwidth]{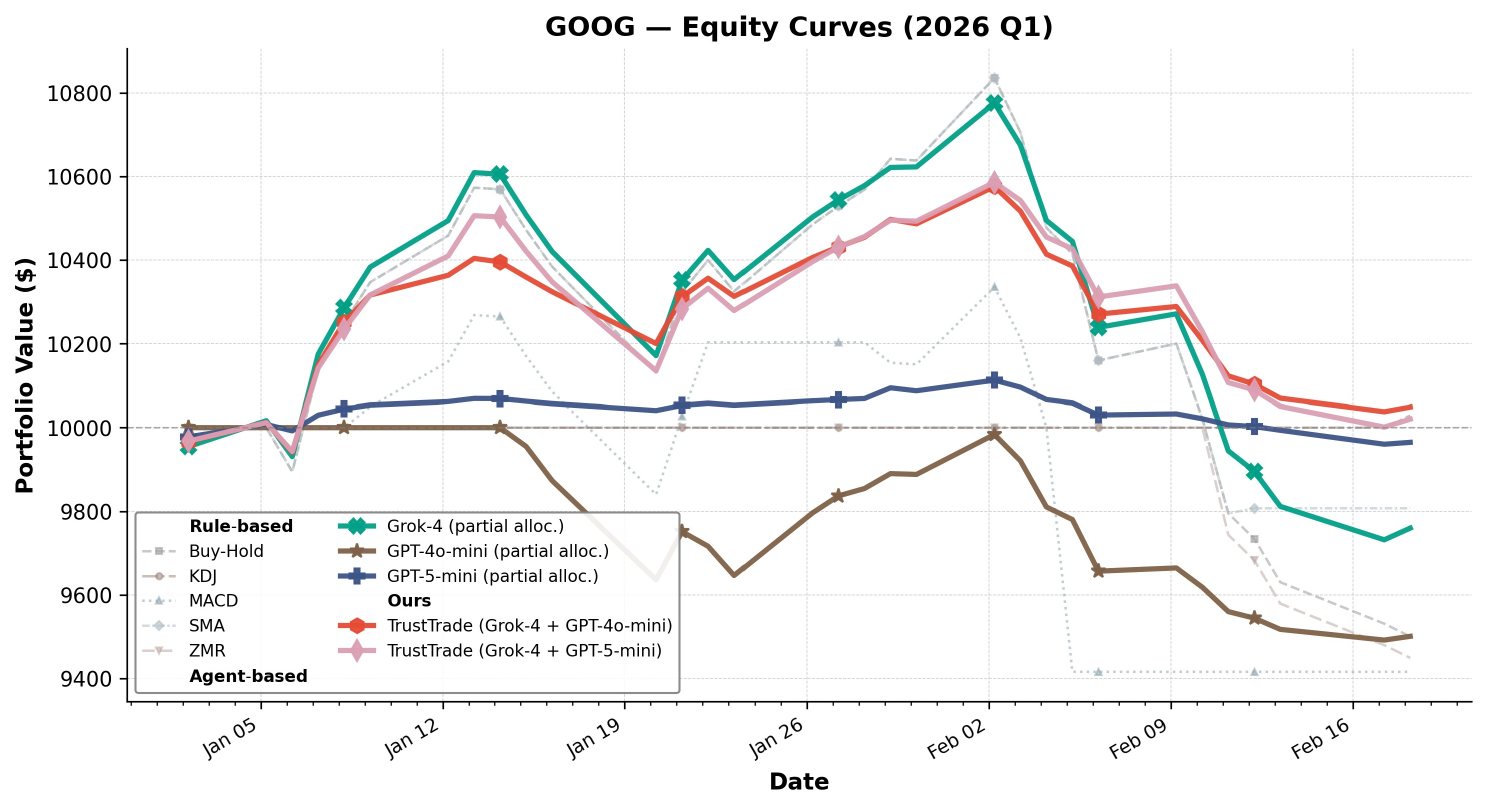}
        \put(-3,48){\textbf{\color{blue} b}}
    \end{overpic}
    \begin{overpic}[width=0.7\textwidth]{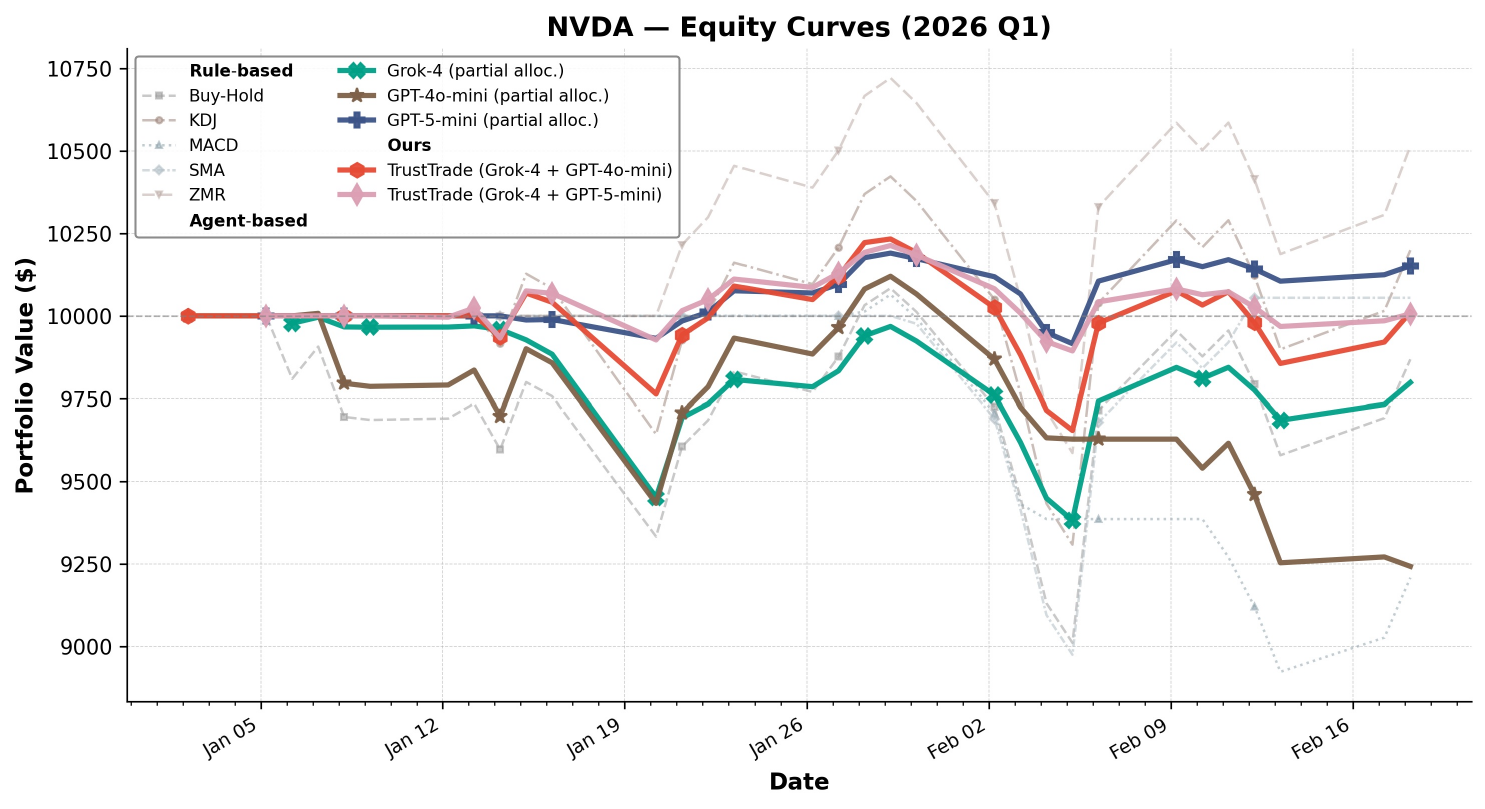}
        \put(-3,48){\textbf{\color{blue} c}}
    \end{overpic}
    \caption{\footnotesize
    \textbf{Daily real-time trading performance on AAPL, GOOG, and NVDA during 2026 Q1.}
    This figure reports day-by-day results in a forward-time setting to reduce potential leakage from earlier market-period evaluation. Rule-based baselines show return swings that closely track price volatility, while pure LLM traders exhibit comparatively unstable behaviors and outcomes. By integrating multi-agent information with selective consensus, \model improves return performance with more stable trading trajectories.
    }
    \label{fig:main_2026q1}
\end{figure}

\subsection*{Study III: Aligning LLM Trading Behavior via Consensus, Temporal Signals and Memory}

Motivated by the behavioural gap identified in Study~II, we next introduce interventions that explicitly target stage-wise stabilization and human-aligned risk control. Fig.~\ref{fig:study3} summarizes how our design choices improve both the \emph{stability} and the \emph{human alignment} of LLM trading. We decompose the effect into three components: selective consensus, deterministic temporal signals, and a memory bank with long/short-term decision reflection, and report their individual contributions to convergence and risk--return behaviour.

\noindent\textbf{TrustTrade overview.}
Fig.~\ref{fig:study3}~{\color{blue}\textbf{a}} provides an overview of TrustTrade, which couples an orchestration layer with a memory bank updated via long/short reflection. For each information domain (fundamentals, market data, news, sentiment), we query multiple heterogeneous LLM agents (e.g., GPT-family and Gemini-family models) to collect independent evidence and form cross-agent domain reports. A credibility scorer then filters and ranks these reports to retain high-consensus information and suppress noisy or conflicting claims. A researcher consolidates the retained evidence before the trader produces the final action. The reflection module equips the memory bank with explicit strategy-evaluation signals (e.g., temporal return and Sharpe slopes over short and long horizons), allowing past decisions to be scored and summarized before being fed back into subsequent decision-making.

\noindent\textbf{Selective consensus improves decision stability.}
Using diverse LLM agents (e.g., GPT-4o-mini and Grok-4) to generate domain reports without filtering leads to poor decision consistency across stages in Fig.~\ref{fig:study3}~{\color{blue}\textbf{b}}. In contrast, introducing selective consensus substantially improves consistency with the final decision by retaining only high-consensus information: our full pipeline maintains high agreement with the final action across intermediate reasoning stages, whereas baseline LLM traders exhibit larger mid-stage reversals before converging at the final step. This stability gain is accompanied by a marked return improvement in the risk--return plane (Fig.~\ref{fig:study3}~{\color{blue}\textbf{c}}), with average cumulative return increasing from roughly $10\%$ to $\sim 26\%$ under the high-consensus configuration, at the cost of an unavoidable increase in MDD from about $3\%$ to $\sim 8\%$.

\noindent\textbf{Temporal signals provide deterministic grounding.}
Fig.~\ref{fig:study3}~{\color{blue}\textbf{d}} shows a temporal-signal summary table generated by our newly introduced analyst that reports deterministic, price-derived trends and indicators. Fig.~\ref{fig:study3}~{\color{blue}\textbf{b}} further shows that introducing these reproducible temporal signals substantially improves decision consistency, bringing LLM decisions better than human stage-wise convergence. At the portfolio level (Fig.~\ref{fig:study3}~{\color{blue}\textbf{c}}), adding temporal signals increases average cumulative return by about $1\%$ while slightly reducing risk (MDD), confirming that deterministic temporal grounding is critical for stabilizing trading decisions.

\noindent\textbf{Memory bank regularizes sequential updates via long/short-term decision reflection.}
Fig.~\ref{fig:study3}~{\color{blue}\textbf{a}} outlines the memory bank with long/short reflection and its role in our orchestration layer. The reflection module evaluates whether a strategy is improving or degrading by tracking temporal performance slopes (e.g., return and Sharpe slopes over short and long horizons) and uses these signals to update the memory bank at test time. This feedback is immediately reused in subsequent trading steps, enabling test-time adaptation: the agent adjusts its allocation and risk posture by borrowing from similar past situations and down-weighting strategies whose recent slopes deteriorate. Compared with variants without memory and reflection, this component substantially reduces risk, with an acceptable trade-off of slightly lower returns, yielding more conservative and robust decisions and shifting the agent configuration toward the human-aligned preference region in Fig.~\ref{fig:study3}~{\color{blue}\textbf{c}}.

\subsection*{Study IV: Overall Benchmark Comparison}

Fig.~\ref{fig:main} and Fig.~\ref{fig:main_2026q1} consolidate all baselines into a unified return--risk comparison. Fig.~\ref{fig:main} involves human annotators as a behavioural reference to contextualize where LLM agents and our method operate on the performance frontier. The plot shows average CR or SR versus average MDD, aggregated across NVDA, AAPL and GOOG. Each point corresponds to one agent configuration; lower MDD or higher SR indicate lower risk, and higher CR indicates higher return. Human annotators (partial allocation) define a human-aligned preference region (shaded ellipse), providing a behavioural reference for acceptable risk exposure.

\noindent\textbf{Risk-return comparison on 2024-Q1.}
%
%
Fig.~\ref{fig:main}~{\color{blue} \textbf{a,b} } summarizes the overall benchmark comparison in 2024-Q1 and highlights three findings. High-return baselines (e.g., buy-and-hold and full-allocation LLM traders) largely lie in a high-drawdown regime, underscoring that raw return alone is misaligned with human risk preferences. Across LLM baselines, model capacity and allocation policy jointly shape a clear return--risk trade-off: LLM agents tend to operate in either a high-return/high-risk regime or a low-return/low-risk regime, with stronger models under full allocation attaining the largest gains but also the largest drawdowns, whereas partial allocation reduces risk at the expense of return, with weaker models clustering near the low-return end. Against this backdrop, our \model based on Grok-4 and GPT-4o-mini model shifts the partial-allocation frontier by improving return at comparable drawdown (and/or reducing drawdown at comparable return), and adding memory and reflection further robustifies performance by lowering drawdown with only a modest return sacrifice, moving the operating point toward the human-aligned region.

\noindent\textbf{Real-time backtesting to mitigate data leakage.}
When trading on historical market periods, a potential concern is data leakage: language models may implicitly encode future price movements through pretraining or contamination, which could artificially inflate performance and distort decision rationales. To address this risk, we conduct a real-time backtest from January~1,2026 to February~18,2026. Each trading day at 1:00~PM, the agent collects contemporaneous information available up to that timestamp, generates a trading decision, and executes the simulated trade, ensuring that future prices are not accessible at decision time.
Fig.~\ref{fig:main_2026q1} plots the day-by-day return curves of all compared methods on AAPL, GOOG, and NVDA in 2026 Q1, while Fig.~\ref{fig:main}~{\color{blue}\textbf{c,d}} reports the corresponding stock-averaged risk--return trade-off comparison. Overall, rule-based strategies fluctuate strongly with market price swings: KDJ achieves a relatively strong return, whereas most other rule-based baselines remain in negative-return regimes. Single-LLM agents are also unstable across assets, with different models peaking on different stocks (e.g., GPT-4o-mini on AAPL, Grok-4 on GOOG, and GPT-5-mini on NVDA). In contrast, our multi-LLM consensus model delivers more stable performance across stocks, reducing risk while improving returns.

\section*{Discussion}
\label{sec:discussion}

We investigated why LLM trading agents often fail to exhibit human-like risk control and decision stability, and proposed \model to align LLM trading behavior with human annotators through three complementary components: (i) selective, credibility-scored multi-agent consensus, (ii) deterministic temporal signals that summarize market state, and (iii) a memory bank updated via long/short reflection for test-time adaptation. Across controlled studies in 2024-Q1, our analysis shows substantial heterogeneity across agents and reasoning pipelines, and reveals that instability is driven by noisy multi-source integration and weak commitment to intermediate evidence. In Study~III, selective consensus improves stage-wise consistency with final actions, temporal signals provide reproducible grounding to reduce hallucination, and memory+reflection further regularize sequential updates by using temporal return/Sharpe slopes to evaluate and adapt strategies.

Overall benchmark results (Study~IV) show that \model improves the return--risk trade-off relative to strong partial-allocation baselines, and that memory with reflection yields a more conservative configuration by reducing drawdown at the cost of a small return decrease. To mitigate concerns about data leakage when evaluating on historical periods, we further introduce a real-time backtesting protocol (2026-Q1) that restricts each decision to contemporaneous information available at that timestamp.

\noindent\textbf{Limitations.}
First, while deterministic temporal signals constrain downstream reasoning, the pipeline still relies on LLM-generated domain reports that may contain omissions, framing bias, or subtle errors. Second, the credibility scorer and consensus filtering can suppress minority-but-correct evidence, particularly in fast-changing or low-coverage market regimes. Third, to enable a controlled comparison to human annotator behavior, we focus on a small set of representative stocks to make the task quantifiable; due to limited API budgets, we also restrict the evaluation to limited time windows. Future evaluations should extend to diversified multi-stock portfolios, which better reflect real-world trading settings and may change the observed return--risk trade-offs. 

\noindent\textbf{Future work.}
Promising directions include (i) real-time backtesting on a broader universe of stocks, including non-megacap and non-technology names, (ii) stronger anti-leakage evaluation protocols (e.g., strict time-stamped data feeds and controlled news snapshots), (iii) learning calibrated confidence and position sizing policies that explicitly target human-aligned drawdown constraints, and (iv) adaptive consensus mechanisms that incorporate uncertainty estimates to avoid discarding rare but informative signals. More broadly, integrating causal market features, transaction costs, and liquidity constraints, and extending the memory bank to support longer-horizon planning and regime detection, could further improve robustness and real-world applicability.
\section*{Methods}
\label{sec:methods}

This section is organized into three parts. We first describe the human trading simulation protocol and data collection pipeline. We then present \model, which integrates selective multi-agent consensus, deterministic temporal-signal analytics, and a reflective memory bank for test-time adaptation. Finally, we summarize experimental details, including rule-based baselines, evaluation metrics, and the human-aligned preference region used for behavioral comparison.

\subsection*{Human Trading Simulation and Result Collection}




To study human decision-making under realistic market information, we designed a stepwise trading simulation with structured logging and persistent storage. The pipeline includes offline data preparation followed by an online interaction loop.

Before user interaction, we preprocess and standardize the raw market, fundamental, and news/sentiment data required by the simulation. To avoid priming participants, we use GPT-4o-mini to collect domain reports as the raw information shown in the interface, and remove any decision-related content (e.g., explicit buy/sell recommendations or position sizes). At runtime, the web interface authenticates participants via a user ID and collects basic demographic/background information (e.g., education level and business/finance experience), then restores prior progress if available or starts a new session by loading the prepared data. Participants iterate through trading days until completion, with automatic portfolio updates and cloud saves after each day.
Each trading day is decomposed into six ordered stages:
{\setlength{\leftmargini}{1.2em}
\begin{list}{}{\setlength{\leftmargin}{1.2em}
               \setlength{\itemsep}{2pt}
               \setlength{\parsep}{0pt}}
\item {d0: Temporal signals} (30-day price history with ticker and indices)
\item {d1: Fundamentals} (e.g., market cap, P/E, EPS)
\item {d2: Market/technical indicators} (e.g., 10/30-day moving averages)
\item {d3: News} (title, source, summary)
\item {d4: Social sentiment} (sentiment and analyst ratings)
\item {Final composite} (aggregated view with execution)
\end{list}
}
At every stage, participants provide an \texttt{action} (BUY/SELL/HOLD), a reliability score (1--100), and a free-text rationale. For stages {d1--d4}, we additionally record a data-leakage flag indicating whether AI-generated decisions were inadvertently visible. In the final stage, participants also specify the most influential source, most reliable source, and trade size (25/50/75/100\%), after which portfolio execution is applied.

To support recovery and analysis, results are written to Amazon Web Services (AWS) S3 under structured keys: full-session exports, portfolio state, and progress markers. This design enables interruption-safe sessions, reproducible behavioral traces, and standardized downstream analysis across participants.

\subsection*{\model: Trust-Rectified Unified Selective Trade}

\begin{figure}[!htbp]
    \centering
    \begin{overpic}[width=0.99\textwidth]{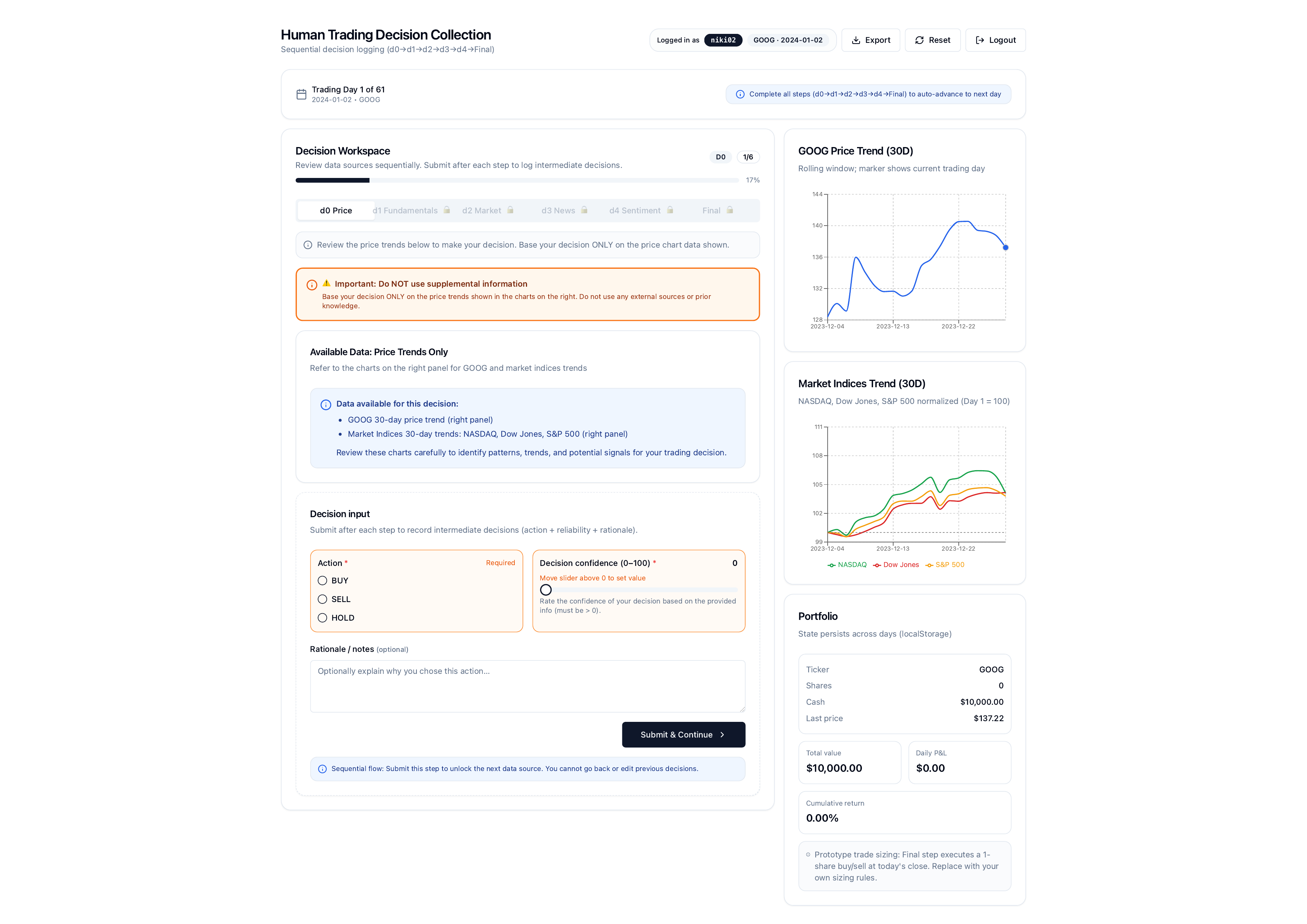}
        \put(8,66){\color{blue} \textbf{a}}
    \end{overpic}
    \caption{\small User interface for the systematic collection of human decision annotations.}
\end{figure}

\begin{figure}[!htbp]
    \centering
    \begin{overpic}[width=0.92\textwidth]{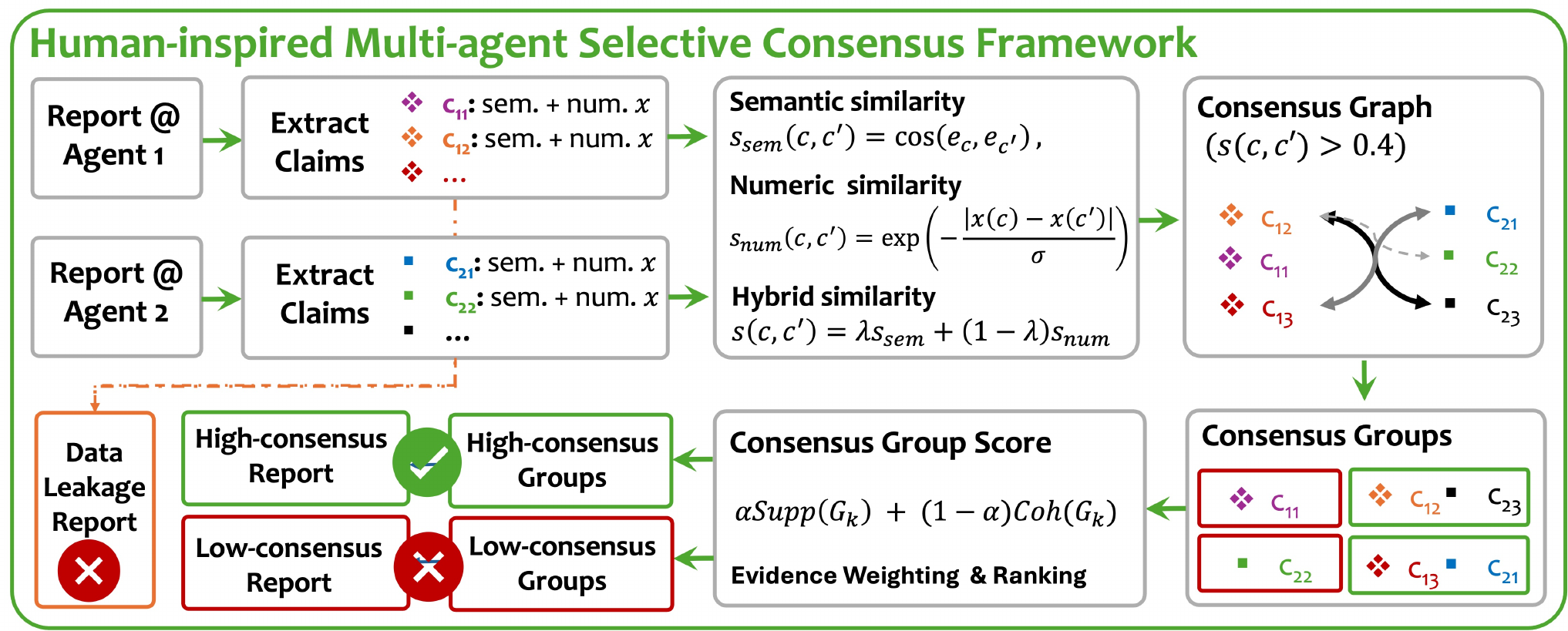}
        \put(-3,38){\color{blue} \textbf{a}}
    \end{overpic}
    \vfill
    \vspace{+0.5em}
    \begin{overpic}[width=0.92\textwidth]{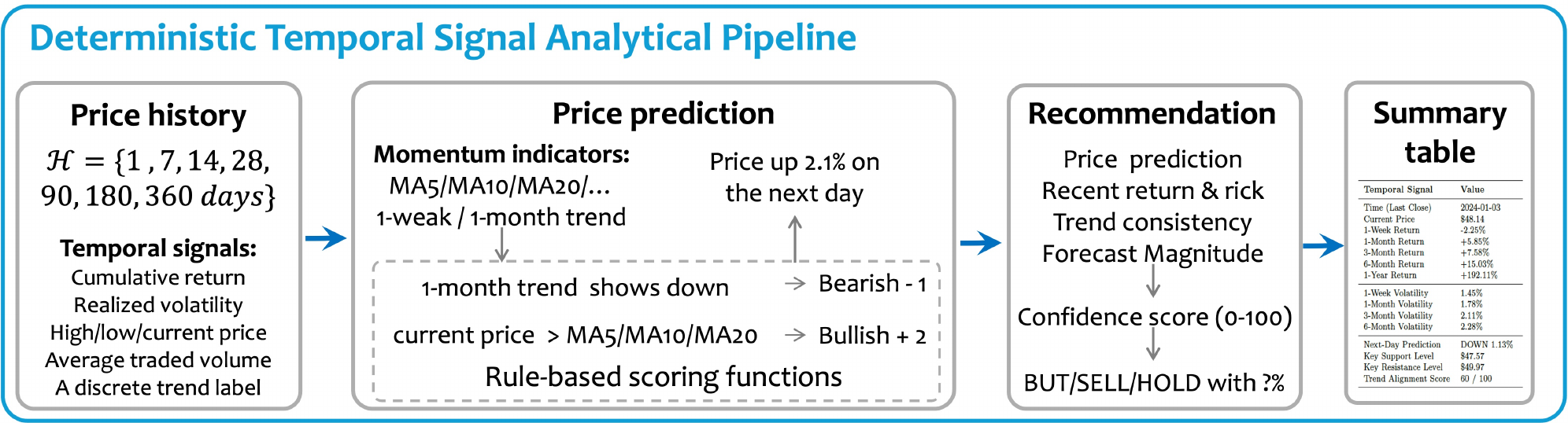}
        \put(-3,25){\color{blue} \textbf{b}}
    \end{overpic}
    \vfill
    \vspace{+0.5em}
    \begin{overpic}[width=0.92\textwidth]{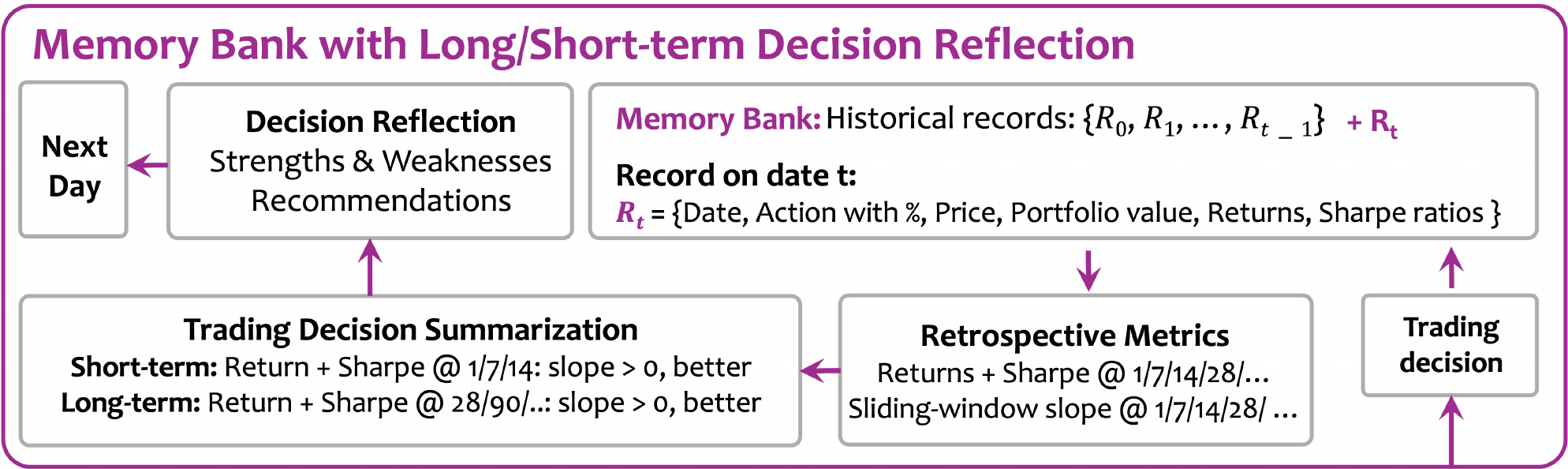}
        \put(-3,26){\color{blue} \textbf{c}}
    \end{overpic}
    \caption{
        \textbf{Three main modules of the porposed \model pipeline.} \textbf{\color{blue} a,} Multi-agent selective consensus for domain report collection and credibility scoring. \textbf{\color{blue} b,} Deterministic price-based analysis module that produces temporal signals and rule-based forecasts. \textbf{\color{blue} c,} Memory bank with short/long-term decision reflection agents that enable test-time adaptation.
    }
    \label{fig:method}
\end{figure}

\noindent\textbf{Overview.} \model, adapted from \cite{xiao2025tradingagents}, comprises three components: (i) multi-agent domain report collection and credibility-scored selective consensus, (ii) deterministic temporal signals with price-derived indicators, and (iii) a memory bank with long/short reflection that supports test-time adaptation.

\noindent\textbf{Human-inspired multi-agent selective consensus framework.}
We introduce a selective consensus mechanism to quantify the reliability of retrieved financial data. The framework departs from the single-agent paradigm by using $N$ independent agents
to collect and parse domain reports in parallel. We then apply domain-wise credibility scoring (fundamentals, market, news, and sentiment) to aggregate these reports: the scorer outputs a structured evidence summary (market signal, source, cross-report consistency, and confidence level), promotes cross-report consistent facts and signals to a high-confidence set, flags conflicting or weakly supported claims as low confidence, and performs a data-leakage audit. The high-confidence summary is used as the default evidence for downstream decision making.

To make credibility scoring more reproducible, we additionally implement a deterministic consensus procedure over atomic claims. For each report $r_i$, we extract a set of point-by-point claims $C_i=\{c_{ij} | j=1, \dots, m_i\}$, where each claim is normalized into a fixed schema (domain, subject, predicate, polarity, optional numeric value, and timestamp). We embed each claim into a vector $e_{c_{ij}}\in\mathbb{R}^d$ and compute the \emph{semantic} similarity between claims from different reports using cosine similarity, denoted by $s_{\mathrm{sem}}(c,c')$. For claims with numeric values, we define a numeric agreement term $s_{\mathrm{num}}(c,c')$. These terms are combined into a final hybrid similarity score:
\begin{equation}
\begin{gathered}
s_{\mathrm{sem}}(c,c') = cos(e_c, e_{c'}),\quad s_{\mathrm{num}}(c,c')=\exp(-|x(c)-x(c')|/\sigma); \\
s(c,c')=\lambda s_{\mathrm{sem}}(c,c')+(1-\lambda)s_{\mathrm{num}}(c,c').
\end{gathered}
\end{equation}
We build a cross-report similarity graph whose nodes are claims and whose edges connect claims from different reports when $s(c,c')\ge\tau$. Connected components define consensus groups $G_k$. For each group, we compute the support ratio and define the cohesion weight as the within-group average pairwise similarity,
\begin{equation}
    \small
    \mathrm{supp}(G_k)=\frac{|\{j:\exists c\in C_j\cap G_k\}|}{N},\quad
    \mathrm{coh}(G_k)
    =
    \frac{2}{N_k(N_k-1)}
    \sum_{\substack{c_u,c_v\in G_k\\ u<v}}
    s(c_u,c_v). 
\end{equation}
where $N_k = |G_k|$ is the number of claims in group $G_k$.
The consensus score is assigned as
$
    S(G_k)=\alpha\,\mathrm{supp}(G_k)+(1-\alpha)\,\mathrm{coh}(G_k). 
$
Groups with high consensus score are retained as high-confidence evidence, while low-support or low-cohesion groups are treated as low-consensus evidence and down-weighted.

\noindent\textbf{Deterministic temporal signal analytical pipeline.}
We implement a deterministic, price-only analytical module that transforms historical finiancial data into four structured outputs: 
{price history}, {price prediction}, {trading recommendations}, and {a summary table}. 
Given a ticker symbol and trade date, the system retrieves daily price data over a maximum lookback window determined by predefined horizons ($\mathcal{H}=\{1,7,14,28,90,180,360\ \text{days}\}$).
The resulting series is standardized into a date-indexed tabular format.

\texttt{Price history.}
For each horizon $h$, the module computes a compact set of descriptive statistics, including cumulative return and realized volatility, high/low/current price levels, average traded volume, and a discrete trend label. Trend is estimated by fitting a polynomial regression to stock prices; the fitted start-to-end change and terminal slope are combined into a single scalar trend score (capturing both direction and strength). We then discretize this score into one of five labels by thresholding its magnitude: large positive values indicate a strong uptrend, small positive values indicate an uptrend, values near zero indicate sideways movement, small negative values indicate a downtrend, and large negative values indicate a strong downtrend.

\texttt{Price prediction.}
The prediction module extracts short-term momentum indicators (MA5/MA10/MA20 alignment, 5-day rate of change, and recent return volatility) and combines them with cross-horizon trend consistency (1-week and 1-month trends) to produce a deterministic bullish/bearish estimate. To ensure reproducible evaluation, we implement a small set of rule-based scoring functions. For example, if the current price is above MA5/MA10/MA20, we increase the bullish confidence by $+2$; if it is below all three moving averages, we increase the bearish confidence by $+2$. The final bullish and bearish scores are then used to output a next-day direction label (\texttt{up}, \texttt{down}, or \texttt{uncertain}), an expected movement magnitude, and the confidence score itself, together with a brief textual rationale. To reduce unstable extrapolation under high-noise conditions, the predicted magnitude is bounded to a conservative interval.

\texttt{Trading recommendations.}
We construct a confidence score in $[0,100]$ by combining four signals: trend consistency across horizons, next-day forecast confidence, recent return strength, and forecasted magnitude. The final action is selected from $\{\texttt{BUY}, \texttt{SELL}, \texttt{HOLD}\}$, and the position size is determined by pre-defined confidence thresholds together with the current portfolio state (position present versus absent). This rule-based layer yields transparent and reproducible recommendations with explicit rationales.

\texttt{Summary table.}
For downstream decision interfaces, the module compiles a concise summary table containing the recommended action and position size, an aggregate confidence score, the next-day forecast (direction and magnitude), and key horizon returns. The final output concatenates the four components above, balancing interpretability (narrative detail) with operational usability (tabular summary).

\noindent\textbf{Memory bank with long/short-term decision reflection.}
To support interruption-safe evaluation and test-time adaptation, \model maintains a memory bank that logs each executed trade together with performance metrics computed over multiple horizons. This memory is later summarized by decision reflection agents into short- and long-term feedback that conditions subsequent decision-making.

\noindent\texttt{Update memory bank and derived metrics.}
For each trading date \(t\), we store a comprehensive trade record
\begin{equation}
    \mathcal{R}_t=\left\{{a_t,\;p_t,\;q_t,\;P_t^{\text{entry}},\;V_t,\;\{R_{t,h}, v_{t,h}^{(R)}\}_{h\in \mathcal{H}},\;\{\text{SR}_{t,h}, v_{t,h}^{(SR)}\}_{h\in \mathcal{H}}}\right\},
\end{equation}
where \(a_t\in\{\texttt{BUY,HOLD,SELL}\}\), \(p_t\) is trade percentage, \(q_t\) is shares changed, \(P_t^{\text{entry}}\) is entry price, and \(V_t\) is pre-trade portfolio value. $R_{t,h}$ and $\text{SR}_{t,h}$ denote the return and Sharpe ratio at horizon $h$, and $v_{t,h}^{(R)}$ and $v_{t,h}^{(RR)}$ denote rolling-window trend slopes (computed over a recent window of trades) for returns and Sharpe ratio, respectively.

To prevent information leakage, we initialize $R_{t,h}$, $\text{SR}_{t,h}$, and their corresponding slope features as missing/empty at trade time $t$. These metrics are backfilled retrospectively only once the future date $t+h$ is observed; at that point, realized return and Sharpe ratio at horizon $h$ (and their rolling-window slopes) are computed and appended to the stored record. We retrospectively update realized return (PnL) for each prior trade \(i<t\):
\begin{equation}
    \small
    R_{t,h}(\%) = 100 \cdot \frac{\Delta \text{PnL}_{t,h}}{V_t},
    \quad
    \Delta \text{PnL}_{t,h} = (P_{t+h}^{\text{entry}}-P_t^{\text{entry}})\cdot d_t \cdot q_t,    
\end{equation}
where $h \in \mathcal{H}$, \(d_t\in\{+1,-1\}\) denotes buy/sell direction.  
For \(h\ge 7\), we also compute annualized Sharpe from daily portfolio-level returns within the horizon:
\begin{equation}
    \small
    \text{SR}_{t,h}
    = \sqrt{252}\cdot \frac{\mu_{t,h}}{\sigma_{t,h}},
    \quad
    \mu_{t,h} 
    = \frac{1}{h}\sum_{\tau=1}^{h} R_{t-\tau,1},
    \quad
    \sigma_{t,h}
    = \sqrt{\frac{1}{h}\sum_{\tau=1}^{h}\left(R_{t-\tau,1}-\mu_{t,h}\right)^2}.
\end{equation}
To capture trend-of-performance (not single-point performance), we compute rolling-window ($w$) linear slopes over recent trades for each horizon:
\begin{equation}
    v^{(R)}_{t,h} = \text{slope}\big(\{R_{t-\tau,h}\}_{\tau<{w}}\big),\qquad
    v^{(\text{SR})}_{t,h} = \text{slope}\big(\{\text{SR}_{t-\tau,h}\}_{\tau<{w}}\big).
\end{equation}
Positive slopes indicate improving recent performance, whereas negative slopes suggest deterioration. We use these trend signals to judge whether the strategy is getting better or worse over time (for both returns and risk-adjusted returns).

\noindent\texttt{Short-term vs. long-term decision reflection.}
\model generates reflections through an agentic function call, which converts stored trading history into an LLM-readable prompt. The memory bank first converts trade records $\{\mathcal{R}_t\}_t$ into a tabular dataframe that summarizes, for each trade date, the action/position and multi-horizon performance statistics (e.g., $r_{i,h}$ and $\text{SR}_{i,h}$) together with their recent trend slopes (e.g., $s_h^{(r)}$ and $s_h^{(S)}$). Conditioned on a selected horizon configuration, the function assembles a compact prompt containing a dataset overview, per-horizon performance/trend summaries, and explicit reflection instructions.

At each trading date $t$, we instantiate two specialized reflection agents that share this functional interface but differ only in horizon configuration: a short-term agent ($\mathcal{H}_{\text{short}}=\{1,7,14\}$) and a long-term agent ($\mathcal{H}_{\text{long}}=\{28,90,180,360\}$). Formally, letting $\Pi(\cdot)$ denote the generated prompt, we obtain
\begin{equation}
F_t^{\text{short}}=\texttt{LLM}_{\text{short}}\left(\{\mathcal{R}_t\}_t, \Pi(\mathcal{H}_{\text{short}})\right),\qquad
F_t^{\text{long}}=\texttt{LLM}_{\text{long}}\left(\{\mathcal{R}_t\}_t, \Pi(\mathcal{H}_{\text{long}})\right).
\end{equation}
The resulting reflections are stored back into the memory bank $\mathcal{R}_t$ and injected into the next decision cycle as historical performance context (used for confidence/risk/position sizing calibration), enabling test-time adaptation. Role-specific reflection memory (Bull/Bear/Trader) is updated separately using realized outcomes and stored for retrieval-augmented reasoning.

\subsection*{Experimental Details}
\noindent\textbf{Rule-based baselines.}
We compare our propsoed \model against several baselines:
{\small\setlength{\itemsep}{2pt}\setlength{\parskip}{0pt}\setlength{\topsep}{2pt}%
\begin{itemize}
    \item \texttt{Buy and Hold}: Investing equal amounts in all selected stocks and holding them throughout the simulation period~\cite{sutton1998reinforcement}.

    \item \texttt{MACD (Moving Average Convergence Divergence)}: A trend-following momentum strategy that generates buy and sell signals based on the crossover points between the MACD line and signal line~\cite{appel1979macd}.

    \item \texttt{KDJ and RSI (Relative Strength Index)}: A momentum strategy combining KDJ (stochastic oscillator) and RSI (relative strength index) indicators to identify overbought and oversold conditions for trading signals~\cite{wilder1978new}.

    \item \texttt{ZMR (Zero Mean Reversion)}: A mean reversion trading strategy that generates signals based on price deviations from and subsequent reversions to a zero reference line~\cite{brock1998heterogeneous}.

    \item \texttt{SMA (Simple Moving Average)}: A trend-following strategy that generates trading signals based on crossovers between short-term and long-term moving averages~\cite{brock1992technical}.
\end{itemize}}

\noindent\textbf{Performance metrics and human-aligned evaluation.}
\texttt{Evaluation metrics.}
We evaluate \model using standard portfolio metrics that capture profitability, risk, and safety relative to baselines. For clarity, we use a unified notation throughout: $T$ is the total number of trading days; $V_t$ is the portfolio (or asset) value at time $t$ (with $V_0$ the initial value); $R_t := \frac{V_t}{V_{t-1}}-1$ is the simple return at time $t$ and $\bar{R} := \frac{1}{T}\sum_{t=1}^T R_t$ is the average return. We denote the per-period risk-free rate by $R_f$, the benchmark return by $b_t$, the active (excess) return by $a_t := R_t-b_t$, and the per-period loss by $L_t := -R_t$.
We report cumulative return, $\text{CR}=\left(\frac{V_T-V_0}{V_0}\right)\times 100\%$, and annualized return, $\text{AR}=\left(\left(\frac{V_T}{V_0}\right)^{\frac{k}{T}}-1\right)\times 100\%$, where $k$ is the number of trading days per year (e.g., $k=252$). We also report the Sharpe ratio, $\mathrm{SR}=\frac{\bar{R}-R_f}{\sigma}$, with volatility defined as $\sigma=\sqrt{\frac{1}{T-1}\sum_{t=1}^T (R_t-\bar{R})^2}$; if reporting annualized Sharpe, we use $\bar{R}_{\text{ann}}=\bar{R}\cdot k$ and $\sigma_{\text{ann}}=\sigma\sqrt{k}$. Finally, we measure maximum drawdown via the running peak $M_t=\max_{0\le s \le t} V_s$, drawdown $DD_t=\frac{V_t-M_t}{M_t}\le 0$, and $\mathrm{MDD}=\max_{0\le t\le T}\left(\frac{M_t-V_t}{M_t}\right)=-\min_{0\le t\le T}DD_t$.

\noindent\texttt{Human-aligned preference region.}
To assess alignment with human annotator behavior, we define a \emph{human-aligned preference region} centered at the empirical human performance.
Let $(\mu_{\mathrm{MDD}}^{(h)}, \mu_{\mathrm{CR}}^{(h)})$ denote the mean MDD and CR of human annotators
under partial allocation, and let $(\sigma_{\mathrm{MDD}}^{(h)}, \sigma_{\mathrm{CR}}^{(h)})$ denote the
corresponding standard errors across stocks.
We define the preference region as an elliptical level set:
\begin{equation}
\mathcal{E}
=
\left\{
(x, y) \;\middle|\;
\left( \frac{x - \mu_{\mathrm{MDD}}^{(h)}}{\sigma_{\mathrm{MDD}}^{(h)}} \right)^2
+
\left( \frac{y - \mu_{\mathrm{CR}}^{(h)}}{\sigma_{\mathrm{CR}}^{(h)}} \right)^2
\leq c
\right\},
\end{equation}
where $c > 0$ controls the size of the preference region. This formulation characterizes the statistical dispersion of human risk-return outcomes rather than imposing hard thresholds. Agents closer to the center of $\mathcal{E}$ exhibit stronger alignment with typical human annotator behavior, while increasing distance indicates greater deviation in risk exposure, return realization, or both. For visualization, $\mathcal{E}$ is rendered with graded shading to reflect degrees of alignment and serves as a soft behavioral reference rather than a strict decision boundary.
\section*{Data availability}

This study involved human annotators providing trading decisions in an experimental setting. All participants provided informed consent prior to participation.
The links below correspond to the web interfaces used to collect human annotations during the trading simulation process: \url{https://tradingagents-human-aapl.netlify.app/}, \url{https://tradingagents-human-goog.netlify.app/}, \url{https://tradingagents-human-nvda.netlify.app/}.
For the simulation itself, all multi-source information presented to participants (e.g., fundamentals, market signals, news summaries, and sentiment cues) was generated from language-model-produced reports. No human-authored analysis, recommendation, or manual curation was included in the decision-facing information stream.

\section*{Code availability}
The code used for data processing, model implementation, and analysis in this study is available at \url{https://github.com/Harvard-AI-and-Robotics-Lab}.

\bibliography{sn-bibliography}

@article{brock1997rational,
  title={Rational expectations and rational learning in a simple asset pricing model},
  author={Brock, William A and Hommes, Cars H},
  journal={Journal of Economic Dynamics and Control},
  volume={21},
  number={8-9},
  pages={1115--1146},
  year={1997},
  publisher={Elsevier}
}

@article{brock1998heterogeneous,
  title={Heterogeneous beliefs and routes to chaos in a simple asset pricing model},
  author={Brock, William A and Hommes, Cars H},
  journal={Journal of Economic dynamics and Control},
  volume={22},
  number={8-9},
  pages={1235--1274},
  year={1998},
  publisher={Elsevier}
}

@article{wei2022chain,
  title={Chain-of-thought prompting elicits reasoning in large language models},
  author={Wei, Jason and Wang, Xuezhi and Schuurmans, Dale and Bosma, Maarten and Xia, Fei and Chi, Ed and Le, Quoc V and Zhou, Denny and others},
  journal={Advances in Neural Information Processing Systems},
  volume={35},
  pages={24824--24837},
  year={2022}
}

@article{hurst2024gpt,
  title={Gpt-4o system card},
  author={Hurst, Aaron and Lerer, Adam and Goucher, Adam P and Perelman, Adam and Ramesh, Aditya and Clark, Aidan and Ostrow, AJ and Welihinda, Akila and Hayes, Alan and Radford, Alec and others},
  journal={arXiv preprint arXiv:2410.21276},
  year={2024}
}

@article{team2023gemini,
  title={Gemini: a family of highly capable multimodal models},
  author={Team, Gemini and Anil, Rohan and Borgeaud, Sebastian and Alayrac, Jean-Baptiste and Yu, Jiahui and Soricut, Radu and Schalkwyk, Johan and Dai, Andrew M and Hauth, Anja and Millican, Katie and others},
  journal={arXiv preprint arXiv:2312.11805},
  year={2023}
}

@article{liu2024deepseek,
  title={Deepseek-v3 technical report},
  author={Liu, Aixin and Feng, Bei and Xue, Bing and Wang, Bingxuan and Wu, Bochao and Lu, Chengda and Zhao, Chenggang and Deng, Chengqi and Zhang, Chenyu and Ruan, Chong and others},
  journal={arXiv preprint arXiv:2412.19437},
  year={2024}
}

@article{tran2025multi,
  title={Multi-agent collaboration mechanisms: A survey of llms},
  author={Tran, Khanh-Tung and Dao, Dung and Nguyen, Minh-Duong and Pham, Quoc-Viet and O'Sullivan, Barry and Nguyen, Hoang D},
  journal={arXiv preprint arXiv:2501.06322},
  year={2025}
}

@article{swanson2024virtual,
  title={The virtual lab: Ai agents design new sars-cov-2 nanobodies with experimental validation. bioRxiv},
  author={Swanson, Kyle and Wu, Wesley and Nash, L Bulaong and Pak, John E and Zou, James},
  year={2024},
  publisher={Published online November}
}

@article{piatti2024cooperate,
  title={Cooperate or collapse: Emergence of sustainability behaviors in a society of llm agents},
  author={Piatti, Giorgio and Jin, Zhijing and Kleiman-Weiner, Max and Sch{\"o}lkopf, Bernhard and Sachan, Mrinmaya and Mihalcea, Rada},
  journal={CoRR},
  year={2024}
}

@article{lin2025llm,
  title={LLM-based Agents Suffer from Hallucinations: A Survey of Taxonomy, Methods, and Directions},
  author={Lin, Xixun and Ning, Yucheng and Zhang, Jingwen and Dong, Yan and Liu, Yilong and Wu, Yongxuan and Qi, Xiaohua and Sun, Nan and Shang, Yanmin and Wang, Kun and others},
  journal={arXiv preprint arXiv:2509.18970},
  year={2025}
}

@article{mandelbrot1963variation,
  title={The variation of certain speculative prices},
  author={Mandelbrot, Benoit and others},
  journal={Journal of business},
  volume={36},
  number={4},
  pages={394},
  year={1963},
  publisher={Springer}
}

@article{engle1982autoregressive,
  title={Autoregressive conditional heteroscedasticity with estimates of the variance of United Kingdom inflation},
  author={Engle, Robert F},
  journal={Econometrica: Journal of the econometric society},
  pages={987--1007},
  year={1982},
  publisher={JSTOR}
}

@book{wilder1978new,
  title={New Concepts in Technical Trading Systems},
  author={Wilder, J. Welles},
  year={1978},
  publisher={Trend Research},
  address={Greensboro, NC}
}

@article{diebold2014network,
  title={On the network topology of variance decompositions: Measuring the connectedness of financial firms},
  author={Diebold, Francis X and Y{\i}lmaz, Kamil},
  journal={Journal of econometrics},
  volume={182},
  number={1},
  pages={119--134},
  year={2014},
  publisher={Elsevier}
}

@article{brock1992technical,
  title={Simple technical trading rules and the stochastic properties of stock returns},
  author={Brock, William and Lakonishok, Josef and LeBaron, Blake},
  journal={The Journal of Finance},
  volume={47},
  number={5},
  pages={1731--1764},
  year={1992},
  publisher={Wiley}
}

@article{li2025r,
  title={R\&D-Agent-Quant: a multi-agent framework for data-centric factors and model joint optimization},
  author={Li, Yuante and Yang, Xu and Yang, Xiao and Xu, Minrui and Wang, Xisen and Liu, Weiqing and Bian, Jiang},
  journal={arXiv preprint arXiv:2505.15155},
  year={2025}
}

@article{yun2025quantevolve,
  title={QuantEvolve: Automating Quantitative Strategy Discovery through Multi-Agent Evolutionary Framework},
  author={Yun, Junhyeog and Lee, Hyoun Jun and Jeon, Insu},
  journal={arXiv preprint arXiv:2510.18569},
  year={2025}
}

@article{han2026quantaalpha,
  title={QuantaAlpha: An Evolutionary Framework for LLM-Driven Alpha Mining},
  author={Han, Jun and Zhang, Shuo and Li, Wei and Yang, Zhi and Dong, Yifan and Hu, Tu and Yuan, Jialuo and Yu, Xiaomin and Zhu, Yumo and Lou, Fangqi and others},
  journal={arXiv preprint arXiv:2602.07085},
  year={2026}
}

@book{sutton1998reinforcement,
  title={Reinforcement learning: An introduction},
  author={Sutton, Richard S and Barto, Andrew G and others},
  volume={1},
  number={1},
  year={1998},
  publisher={MIT press Cambridge}
}

@article{guo2026meme,
  title={MEME: Modeling the Evolutionary Modes of Financial Markets},
  author={Guo, Taian and Shen, Haiyang and Luo, Junyu and Xing, Zhongshi and Lian, Hanchun and Huang, Jinsheng and Chen, Binqi and Liu, Luchen and Ma, Yun and Zhang, Ming},
  journal={arXiv preprint arXiv:2602.11918},
  year={2026}
}

@article{xiao2025tradingagents,
  title={TradingAgents: Multi-Agents LLM Financial Trading Framework},
  author={Xiao, Yijia and Sun, Edward and Luo, Di and Wang, Wei},
  journal={arXiv preprint arXiv:2412.20138},
  year={2025}
}

@article{zhang2024stockagent,
  title={When AI Meets Finance (StockAgent): Large Language Model-based Stock Trading in Simulated Real-world Environments},
  author={Zhang, Chong and Liu, Xinyi and Zhang, Zhongmou and others},
  journal={arXiv preprint arXiv:2407.18957},
  year={2024}
}

@article{cao2025chainofalpha,
  title={Chain-of-Alpha: Unleashing the Power of Large Language Models for Alpha Mining in Quantitative Trading},
  author={Cao, Lang and Xi, Zekun and Liao, Long and Yang, Ziwei and Cao, Zheng},
  journal={arXiv preprint arXiv:2508.06312},
  year={2025}
}

@article{ding2024survey,
  title={Large Language Model Agent in Financial Trading: A Survey},
  author={Ding, Y and Li, J and Wang, X and Chen, H},
  journal={arXiv preprint arXiv:2408.06361},
  year={2024}
}

@article{lopezlira2025orderbook,
  title={Can Large Language Models Trade? Testing Financial Theories with LLM Agents in Market Simulations},
  author={Lopez-Lira, Alejandro and others},
  journal={arXiv preprint arXiv:2504.10789},
  year={2025}
}

@article{bai2024rlreview,
  title={A Review of Reinforcement Learning in Financial Applications},
  author={Bai, Y and Gao, S and Wan, J and Zhang, L and Song, H},
  journal={arXiv preprint arXiv:2411.12746},
  year={2024}
}

@article{unnik2024sentiment,
  title={Financial News-Driven LLM Reinforcement Learning for Portfolio Management},
  author={Unnikrishnan, Ananya},
  journal={arXiv preprint arXiv:2411.11059},
  year={2024}
}

@article{coriat2025harlf,
  title={HARLF: Hierarchical Reinforcement Learning and Lightweight LLM-Driven Sentiment Integration for Financial Portfolio Optimization},
  author={Coriat, Thomas and Benhamou, Eric},
  journal={arXiv preprint arXiv:2507.18560},
  year={2025}
}

@article{benhenda2025finrldeepseek,
  title={FinRL-DeepSeek: LLM-Infused Risk-Sensitive Reinforcement Learning for Trading Agents},
  author={Benhenda, Elias},
  journal={arXiv preprint arXiv:2502.07393},
  year={2025}
}

@article{lee2025bias,
  title={Your AI, Not Your View: The Bias of LLMs in Investment Analysis},
  author={Lee, Hoyoung and Seo, Junhyuk and Park, Suhwan and Lee, Junhyeong and Ahn, Wonbin and Choi, Chanyeol and Lopez-Lira, Alejandro and Lee, Yongjae},
  journal={arXiv preprint arXiv:2507.20957},
  year={2025}
}

@book{ingersoll1987theory,
  title={Theory of financial decision making},
  author={Ingersoll, Jonathan E},
  volume={3},
  year={1987},
  publisher={Bloomsbury Publishing PLC}
}

@book{kahneman2011thinking,
  title={Thinking, Fast and Slow},
  author={Kahneman, Daniel},
  volume={0},
  year={2011},
  publisher={Farrar, Straus and Giroux}
}

@article{tuckett2011emotions,
  title={Emotions and Financial Markets},
  author={Tuckett, David and Taffler, Richard},
  journal={Handbook of Behavioural Economics and Smart Decision-Making},
  year={2011},
  publisher={Wiley-Blackwell}
}

@article{coval2001information,
  title={Information and Noise in Financial Markets: An Experimental Study},
  author={Coval, Joshua D. and Shumway, Tyler},
  journal={Journal of Finance},
  year={2001},
  volume={56},
  pages={1141--1179}
}

@article{froot1992explaining,
  title={Explaining Investor Beliefs},
  author={Froot, Kenneth and David S. Scharfstein and Jeremy C. Stein},
  journal={Journal of Finance},
  year={1992},
  volume={47},
  pages={lias}
}

@article{wang2024survey,
  title={A survey on large language model based autonomous agents},
  author={Wang, Lei and Ma, Chen and Feng, Xueyang and Zhang, Zeyu and Yang, Hao and Zhang, Jingsen and Chen, Zhiyuan and Tang, Jiakai and Chen, Xu and Lin, Yankai and others},
  journal={Frontiers of Computer Science},
  volume={18},
  number={6},
  pages={186345},
  year={2024},
  publisher={Springer}
}

@article{preis2013quantifying,
  title={Quantifying trading behavior in financial markets using Google Trends},
  author={Preis, Tobias and Moat, Helen Susannah and Stanley, H Eugene},
  journal={Scientific reports},
  volume={3},
  number={1},
  pages={1684},
  year={2013},
  publisher={Nature Publishing Group UK London}
}

@misc{llmtrading2025moerouting,
  author       = {Anonymous},
  title        = {LLM-Based Routing in Mixture of Experts: A Novel Framework for Trading},
  howpublished = {arXiv:2501.09636},
  year         = {2025},
  archivePrefix= {arXiv},
  eprint       = {2501.09636}
}

@misc{llmtrading2024drlportfoliochina,
  author       = {Anonymous},
  title        = {A Deep Reinforcement Learning Framework for Dynamic Portfolio Optimization: Evidence from China's Stock Market},
  howpublished = {arXiv:2412.18563},
  year         = {2024},
  archivePrefix= {arXiv},
  eprint       = {2412.18563}
}

@misc{llmtrading2025decisioninformednn,
  author       = {Anonymous},
  title        = {Decision-informed Neural Networks with Large Language Model Integration for Portfolio Optimization},
  howpublished = {arXiv:2502.00828},
  year         = {2025},
  archivePrefix= {arXiv},
  eprint       = {2502.00828}
}

@misc{llmtrading2024multimodalfoundationagent,
  author       = {Anonymous},
  title        = {A Multimodal Foundation Agent for Financial Trading: Tool-Augmented, Diversified, and Generalist},
  howpublished = {arXiv:2402.18485},
  year         = {2024},
  archivePrefix= {arXiv},
  eprint       = {2402.18485}
}

\end{document}